\documentclass[]{aastex631}

\usepackage{bm}
\usepackage{amsmath}
\usepackage{comment}

\newcommand\ltsima{$\; \buildrel <\over\sim \;$}
\newcommand\simlt{\lower.5ex\hbox{\ltsima}}
\newcommand\gtsima{$\; \buildrel >\over\sim \;$}
\newcommand\simgt{\lower.5ex\hbox{\gtsima}}

\newcommand{\like}{\mathcal{L}}
\newcommand{\prior}{\pi}
\newcommand{\data}{\mathcal{D}}
\newcommand{\gal}{\mathrm{Gal}}

\shorttitle{Bayesian Analysis with Physically Motivated Galactic Priors}
\shortauthors{Nunota and Masuda}

\graphicspath{{./}}

\begin{document}

\title{
A Bayesian Inference Framework for Binary Lens Events Fully Incorporating Higher-Order Effects 
}

\author[0009-0005-3414-455X]{Kansuke Nunota}
\affiliation{Department of Earth and Space Science, Graduate School of Science, Osaka University, Toyonaka, Osaka 560-0043, Japan}

\author[0000-0003-1298-9699]{Kento Masuda}
\affiliation{Department of Earth and Space Science, Graduate School of Science, Osaka University, Toyonaka, Osaka 560-0043, Japan}

\begin{abstract}
Higher-order effects, such as microlensing parallax and lens orbital motion, are essential for characterizing the physical properties of microlensing planetary systems including their masses and distances.
In practice, higher-order effects are often included in light-curve modeling only when their signals are strong, because their inclusion in weakly constrained cases tends to drive the inferred parameters toward regions of parameter space that are disfavored by standard Galactic models.
This binary choice over physically continuous effects introduces an event-dependent selection function that complicates population-level interpretation and prevents strong and weak detections from being combined within a single uniform framework.
In this work, we investigate the origin of instabilities associated with higher-order effects, and show that such instabilities can arise from a mismatch between commonly adopted uninformative priors in the light-curve parameter space and the distributions predicted by Galactic models.
Motivated by this, we develop a new Bayesian framework that enables higher-order effects to be incorporated in a stable and uniform manner.
As part of this framework, we introduce Galactic Prior Modeling Engine (\texttt{gapmoe}), a dedicated tool that enables efficient evaluation of the Galactic prior density and allows it to be incorporated directly into the light-curve inference. 
Using simulated events, we demonstrate that our framework robustly recovers the true physical parameters while fully accounting for microlensing parallax and lens orbital motion. 
This framework eliminates the need for ad-hoc model selection and provides a scalable and statistically consistent pathway for analyzing the large samples expected from next-generation surveys such as the \textit{Roman Space Telescope}.
\end{abstract}


\section{Introduction} \label{sec-intro}
Higher-order effects play a central role in determining the physical properties of microlensing planetary systems. 
Microlensing parallax \citep{gou92} enables a direct measurement of the lens mass and distance \citep[e.g.,][]{mur11,fur13}, while lens orbital motion \citep[LOM;][]{dom98,iok99} provides direct constraints on the orbital configuration of the lens system \citep[e.g.,][]{ben10,shi11,shi12}.

In principle, parallax and LOM are always present and can be incorporated into the light-curve model.
In practice, however, the treatment depends on the strength of the signal.
When their signals are sufficiently strong, they are modeled explicitly, and the resulting light-curve parameters are directly transformed into physical quantities of the lens system.
In contrast, when higher-order effects are only weakly constrained by the data, they are frequently omitted from the light-curve model. In such cases, the physical properties of the lens are instead inferred through a Bayesian analysis based on Galactic models \citep[e.g.,][]{suz14,han25}.
The motivation for excluding higher-order effects in weak-signal cases is that their inclusion often leads to unstable inference: the corresponding parameters can drift toward physically implausible regions, such as anomalously large parallax values \citep[e.g.,][]{rat17}. 
For this reason, the choice of whether to include higher-order effects is typically made on an event-by-event basis.

This ad-hoc strategy is problematic for at least two reasons. 
First, in this procedure, physically continuous effects are treated in a binary manner. 
In practice, this requires introducing a detection threshold, for example in $\Delta\chi^2$, to decide whether higher-order effects are included. Such a threshold is inherently ad hoc, and even a uniform choice does not lead to a uniform selection in the physical parameter space: the probability that an event passes the threshold depends not only on its signal-to-noise ratio, sampling, and residual systematics, but also on its physical properties. Consequently, the selected sample is subject to a complicated event-dependent selection function, making population-level interpretation difficult and preventing a statistically clean combination of strong and weak detections within a single framework.
Second, omitting higher-order effects discards potentially valuable information encoded in the light curve, even when those effects are only weakly detected.

These issues highlight the need for an inference framework in which higher-order effects can be incorporated in a stable and uniform manner, regardless of the strength of their signals. 
This capability is particularly important in the context of upcoming large-scale microlensing surveys, such as the \textit{Roman Space Telescope}, where a statistically homogeneous treatment of higher-order effects across a large event sample will be essential.

In this work, we first clarify the origin of the instabilities associated with weak higher-order effects, along with other limitations of the conventional two-step approach. 
We show in Section~\ref{sec:lim_conv_method} that a key issue lies in the use of weakly informative priors in the light-curve parameter space, which can induce strongly biased and physically implausible prior distributions in the physical parameter space. 
We then address this issue by developing a Bayesian inference framework that incorporates physically motivated Galactic priors directly into the light-curve analysis in Section~\ref{sec:method}. 
Implementing this framework requires efficient evaluation of probability densities defined by Galactic population models, for which we introduce a dedicated tool, Galactic Prior Modeling Engine (\texttt{gapmoe})\footnote{\url{https://github.com/NunotaKansuke/gapmoe}}.
We test the proposed framework using simulated data in Section~\ref{sec:result_new} and evaluate its performance relative to the conventional two-step approach. In Section~\ref{sec:discussion}, we discuss differing perspectives regarding the treatment of higher-order effects, the roles of priors in event- and population-level inference, and the broader applicability of \texttt{gapmoe}. Finally, Section~\ref{sec:summary} summarizes our conclusions and outlines future prospects.

\section{Limitations of the conventional Bayesian Analysis}\label{sec:lim_conv_method}

Conventional light-curve analyses often yield unstable or unphysical inferences when higher-order effects are only weakly constrained by the data.
In this section, we first review the mathematical structure of the conventional ``two-step'' Bayesian inference framework and clarify a key source of this instability.

Below we use $\bm{y}$ to denote the physical parameters of the lens system and the source, which include the lens mass $M_{\rm L}$, the lens and source distances $D_{\rm L}$ and $D_{\rm S}$, and the relative proper motion $\bm{\mu}_{\rm rel}$.
While $\bm{y}$ represents the set of physical parameters of primary interest, microlensing analyses typically adopt an alternative parameterization, denoted by $\bm{x}$, for model fitting. Typical components of $\bm{x}$ include the time of closest approach $t_0$, impact parameter $u_0$, event timescale $t_{\rm E}$, mass ratio $q$, projected separation $s$, and trajectory angle $\alpha$. Depending on the model complexity, $\bm{x}$ may also include parameters describing higher-order effects, such as the finite-source parameter $\rho$, the microlensing parallax vector $\bm{\pi}_{\rm E}$, and lens-orbital-motion (LOM) parameters $\bm{\gamma}$. In this sense, the use of $\bm{x}$ can be viewed as a reparameterization of the inference problem, replacing $\bm{y}$ with a set of parameters more directly tied to the light-curve morphology.
In general, the mapping from the physical parameters to the light-curve parameters, $\bm{x}=f(\bm{y})$, is deterministically specified by the microlensing geometry. The inverse transformation, $\bm{y}=f^{-1}(\bm{x})$, however, is not generally well defined; its uniqueness depends on whether higher-order effects are included in $\bm{x}$.

\subsection{Conventional ``Two-step'' Inference Framework}\label{sec:conv_method}

In a standard analysis of a microlensing light curve, the information on the set of model parameters $\bm{x}$ is summarized in the form of samples from the posterior probability density function (PDF) $p_1 (\bm{x}|\data)$, given the light-curve data $\data$ and the prior PDF $\prior_1(\bm{x})$:
\begin{align}
\label{eq:post_x}
    p_1 (\bm{x}|\data) \propto \like (\bm{x}) \cdot \prior_1(\bm{x}).
\end{align}
Here the likelihood function $\like(\bm{x}) \equiv p(\data|\bm{x})$ is defined as the probability to obtain data given $\bm{x}$. 
For example, if the error $\sigma_i$ in the observed flux $f_i$ follows independent Gaussian distributions,
\begin{align}
\label{eq:likelihood_normal}
    \like(\bm{x}) = \prod_i {1 \over \sqrt{2\pi\sigma_i^2}}\,\exp\left[-{1\over 2}\left( f_i - m_i(\bm{x}) \over \sigma_i\right)^2\right],
\end{align}
where $m_i(\bm{x})$ denotes the model-predicted flux at the epoch of the $i$th data point.\footnote{More flexible likelihood models that include correlated photometric noise are discussed in Section~\ref{sec:future_prospects}.}
Typically, the prior PDF $\pi_1(\bm{x})$ is chosen to be separable and weakly informative (e.g., uniform), and the sampling is performed using Markov Chain Monte Carlo (MCMC) methods.

When higher-order effects are weak and therefore excluded from the light-curve model, the light-curve parameter set $\bm{x}$ does not uniquely determine the physical parameters $\bm{y}$.
In practice, this situation is typically treated using what is commonly referred to as a ``Bayesian analysis''
to infer $\bm{y}$ by combining the light-curve data $\data$ with a prior usually (but not necessarily) informed by Galactic models.\footnote{
More generally, Bayesian inference is applicable even when the mapping from $\bm{x}$ to $\bm{y}$ is unique; the distinction is just that the role of the prior becomes particularly prominent when this mapping is degenerate.
Moreover, even when one samples $\bm{x}$ under the assumption of a uniform prior and then transforms uniquely to $\bm{y}$, this procedure still constitutes a Bayesian analysis; here the prior on $\bm{x}$ induces a nontrivial prior PDF on $\bm{y}$.
This observation is closely related to the key themes of this paper.
}
The goal of this analysis is to obtain samples from the posterior PDF for a set of physical parameters $\bm{y}$ given the light-curve data $\data$ and the prior PDF for $\bm{y}$, $\prior_\gal(\bm{y})$:
\begin{align}
\label{eq:post_y}
    p_\gal (\bm{y}|\data) \propto p(\data|\bm{y}) \cdot \prior_\gal(\bm{y}).
\end{align}
A commonly used method to 
sample from Equation~\ref{eq:post_y} is to draw samples of the physical parameters $\bm{y}$ from the Galactic prior $\prior_\gal(\bm{y})$ and assign each sample a weight given by the likelihood $p(\data|\bm{y})=\like(\bm{x}=f(\bm{y}))$. 
The likelihood evaluation is performed by recycling the samples from $p_1(\bm{x}|\data)$ conditioned on an uninformative prior (Equation~\ref{eq:post_x}): knowing $\pi_1(\bm{x})$ used for this sampling, the posterior samples can be translated into the unnormalized likelihood function
\begin{align}
\label{eq:likelihood_from_p1}
    \like(\bm{x}) \propto {p_1 (\bm{x}|\data) \over \prior_1(\bm{x})},
\end{align}
where $p_1(\bm{x}|\data)$ is usually approximated by an independent Gaussian based on the samples, and $\like(\bm{x})$ is evaluated at $\bm{x}=f(\bm{y})$. Note that the prior depedence must be explictly corrected.

In what follows, we refer to this procedure as the two-step approach: (i) the posterior $p_1(\bm{x}\mid\data)$ is first inferred under an uninformative prior in the light-curve parameter space, and (ii) an approximate likelihood constructed from these samples is then used to reweight samples drawn from the Galactic prior to obtain $p_\gal(\bm{y}\mid\data)$.

\begin{deluxetable*}{cc|cc|c}
\tablecaption{Summary statistics for Event~1: central 90\% intervals (p5--p95), true values, and prior distributions.\label{table:event2_params}}
\tablehead{
\colhead{Parameter} & \colhead{Unit} & \colhead{$90\%$ interval} & \colhead{Prior} & \colhead{True}
}
\startdata
\cutinhead{Light-curve parameters}
$t_0 - 10085$ & day & $0.003\,\text{--}\,0.039$ & $\mathcal{U}(-10,10)$ & $0.000$ \\
$t_{\rm E}$ & day & $61.67\,\text{--}\,69.58$ & $\mathcal{U}(60,80)$ & $68.51$ \\
$u_0$ & $10^{-2}$ & $0.99\,\text{--}\,1.13$ & $\mathcal{U}(-500,500)$ & $1.00$ \\
$\rho$ & $10^{-3}$ & $0.08\,\text{--}\,0.11$ & $\mathcal{U}(0,200)$ & $0.09$ \\
$q$ & $\log_{10}$ & $-2.32\,\text{--}\,-2.25$ & $\mathcal{U}(-5,0)$ & $-2.30$\\
$s$ & --- & $0.948\,\text{--}\,0.953$ & $\mathcal{U}(0,5)$ & $0.950$ \\
$\alpha$ & rad & $2.851\,\text{--}\,2.883$ & $\mathcal{U}(-\pi,\pi)$ & $2.867$ \\
$\pi_{\rm E,N}$ & $10^{-1}$ & $-5.01\,\text{--}\,3.08$ & $\mathcal{U}(-10,10)$ & $1.29$ \\
$\pi_{\rm E,E}$ & $10^{-1}$ & $0.24\,\text{--}\,4.12$ & $\mathcal{U}(-10,10)$ & $1.36$ \\
\cutinhead{Physical parameters}
$M_{\rm L}$ & $M_\odot$ & $0.13\,\text{--}\,0.72$ & --- & $0.52$ \\
$D_{\rm L}$ & kpc & $1.79\,\text{--}\,5.07$ & --- & $3.66$ \\
$\mu_{\rm rel,N}$ & mas\,yr$^{-1}$ & $-3.17\,\text{--}\,3.88$ & --- & $3.00$ \\
$\mu_{\rm rel,E}$ & mas\,yr$^{-1}$ & $0.39\,\text{--}\,3.66$ & --- & $3.00$ \\
\enddata
\end{deluxetable*}
\begin{figure}[t]
    \centering
    \includegraphics[width=0.85\textwidth]{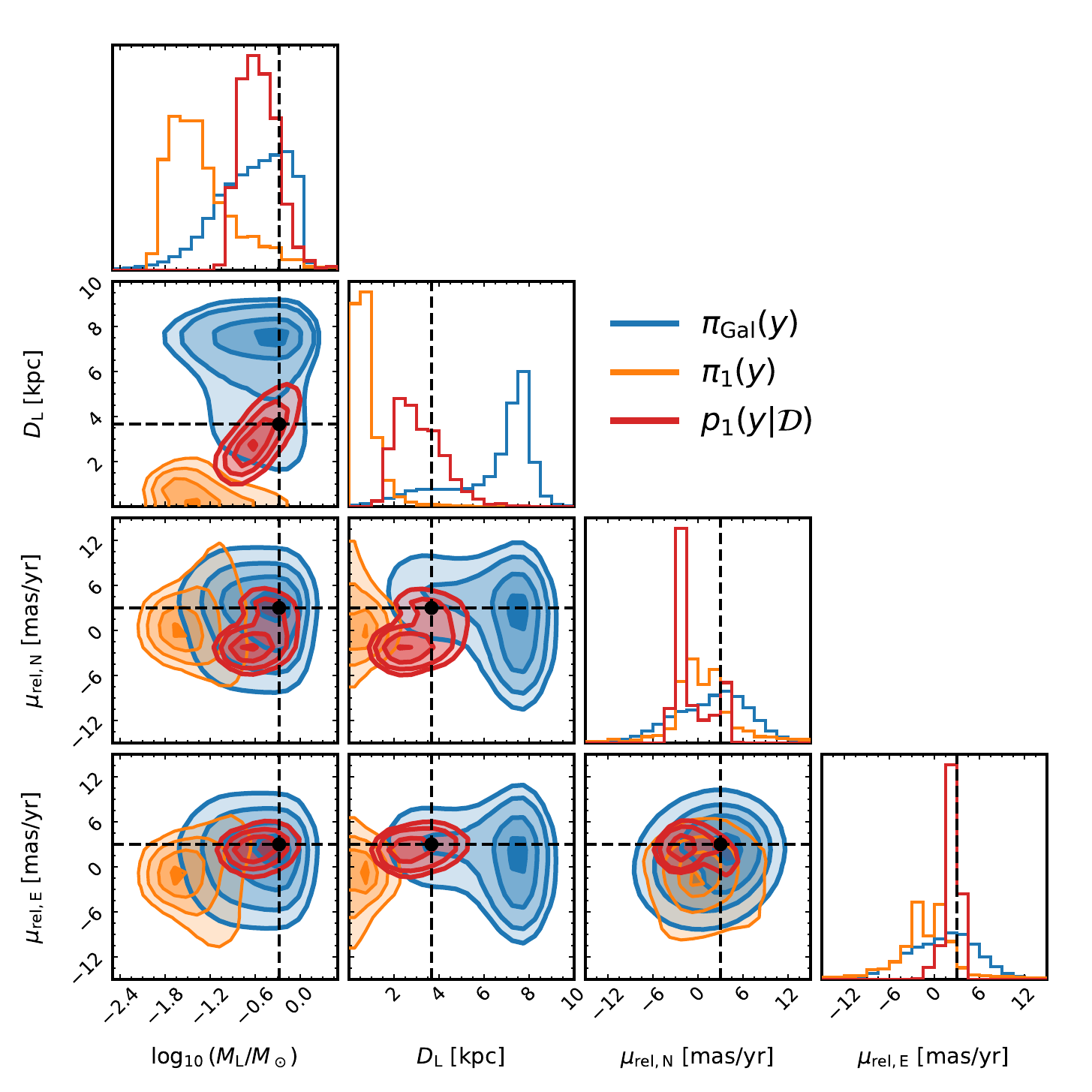}
    \caption{
    Joint distributions in the physical parameter space for Event~1.
    Here, $\prior_\gal(\bm{y})$ denotes the Galactic-model prior, $\pi_1(\bm{y})$ the physical-space prior induced by the weakly informative light-curve prior $\pi_1(\bm{x})$, and $p_1(\bm{y}\mid\data)$ the transformed posterior obtained from the posterior $p_1(\bm{x}\mid\data)\propto\mathcal{L}(\bm{x})\pi_1(\bm{x})$.
    The blue, orange, and red contours show these three distributions, respectively.
    The black dashed lines and black markers indicate the true input parameters.
        }
    \label{fig:prior_comp}
\end{figure}

\begin{figure}[t]
    \centering
    \includegraphics[width=0.49\textwidth]{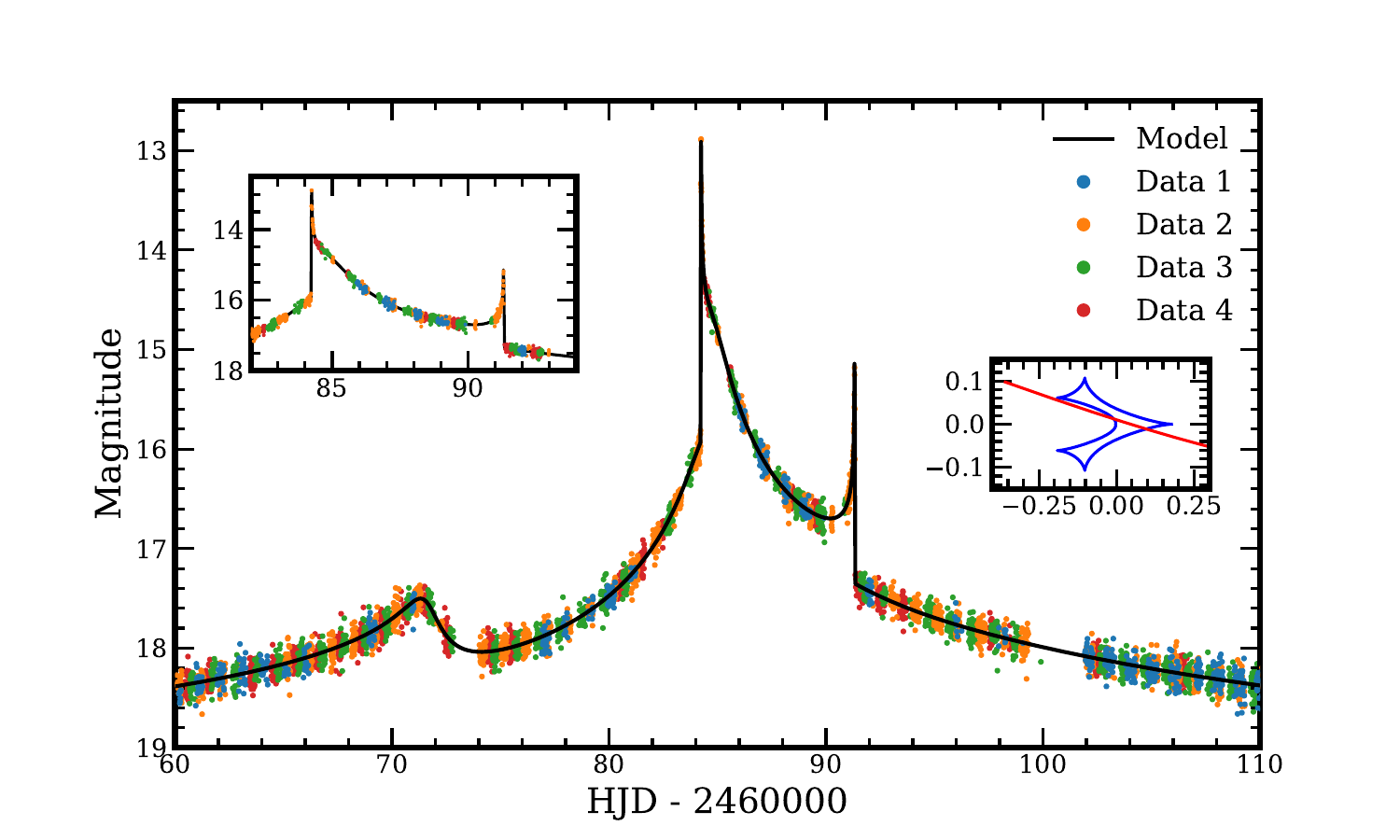}
    \includegraphics[width=0.49\textwidth]{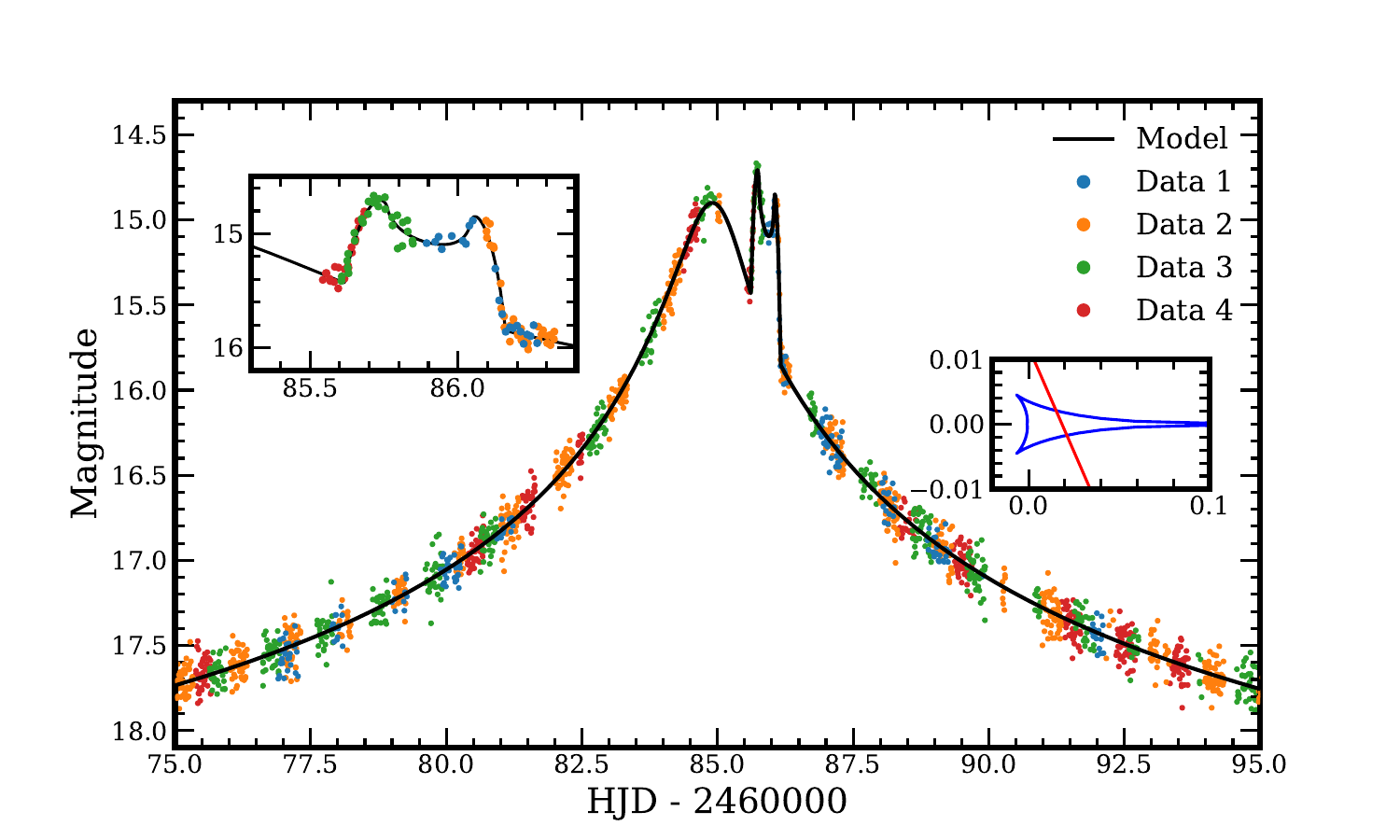}
    \caption{
        Simulated $I$-band microlensing light curves for Event~1 (left) and Event~2 (right).
        Event~1 represents a typical event with a strong parallax signal, whereas Event~2 represents a typical event with weak higher-order signals.
        The main panels show the multi-band light curves and the input models.
        Data~1--4 correspond to the four observing sites.
        Insets highlight the anomaly regions and the caustic structures.
    }
    \label{fig:simu_lc}
\end{figure}

\subsection{Bias Induced by ``Non-informative'' Prior} \label{sec:noninformative_prior_bias}

In principle, the light-curve parameter vector $\bm{x}$ should include higher-order effects because such effects are physically always present.
In practice, however, the corresponding parameters are often excluded when their signals are weak, since including them can lead to unstable inferences in which the parameters drift toward physically implausible regions.
For example, fits including microlensing parallax may yield anomalously large values of the parallax vector $(\pi_{\mathrm{E},N},\pi_{\mathrm{E},E})$ \citep[e.g.,][]{rat17}.

Such behavior is often attributed to photometric systematics being absorbed as higher-order microlensing signals \citep[e.g.,][]{kos14}.
In particular, the formal uncertainties reported by difference image analysis (DIA; \citealt{tom96, ala98}) may underestimate the true photometric scatter because they do not fully capture residual image-subtraction artifacts, long-term trends, observing-condition-dependent variations, or time-correlated noise.
Consequently, residual structures that are unrelated to microlensing may be interpreted by the model as weak parallax or lens-orbital-motion signals.
In terms of Equation~\ref{eq:likelihood_normal}, this corresponds to a misspecified likelihood: the model $m(\bm{x})$ may not describe all flux variations present in the data, and the residuals $f-m$ may neither be independent nor Gaussian with the reported uncertainties.
Such effects can be treated more flexibly by extending the likelihood to include correlated noise, for example through Gaussian-process models, as discussed in Section~\ref{sec:future_prospects}.

However, likelihood misspecification does not necessarily account for the full origin of unstable or unphysical higher-order solutions.
A second source of bias can arise from the prior adopted in the light-curve parameter space, even when the likelihood is idealized and correctly specified.
We examine this effect below using a commonly adopted form of direct physical-parameter inference.

When higher-order effects such as microlensing parallax are included in the light-curve model, it is common to adopt weakly informative, often uniform, priors on the light-curve parameters $\bm{x}$ and then transform the resulting samples into physical parameters $\bm{y}$.
As a representative example of such a prior choice, we consider the uniform priors on $\bm{x}$ listed in Table~\ref{table:event2_params}, which are also adopted in the simulation test below.
We draw samples from these priors and transform them into the corresponding physical parameters using $(t_{\rm E}, \rho, \pi_{\rm E,N}, \pi_{\rm E,E})$ (see Appendix~\ref{sec:parameterization} for details).
The transformation to the lens distance $D_{\rm L}$ additionally requires the source distance $D_{\rm S}$.
Here we draw $D_{\rm S}$ from a prior distribution generated with \texttt{genulens}\footnote{\url{https://github.com/nkoshimoto/genulens}} \citep{kos21,kos22}.
This treatment of $D_{\rm S}$, or the simpler approximation of fixing $D_{\rm S}\simeq 8~\mathrm{kpc}$, is commonly used when the microlensing parallax signal is regarded as detected \citep[e.g.,][]{mur11,fur13}.

This procedure induces a prior distribution $\pi_1(\bm{y})$ in the physical parameter space. For a mapping $\bm{x}=f(\bm{y})$ from the physical parameters to the light-curve parameters,\footnote{
Strictly speaking, for the transformation considered here, we augment the parameter vector with $D_{\rm S}$ and take
$\bm{x}=(t_{\rm E},\rho,\pi_{\rm E,N},\pi_{\rm E,E},D_{\rm S})$ and $\bm{y}=(M_{\rm L},D_{\rm L},\mu_{\rm rel,N},\mu_{\rm rel,E},D_{\rm S})$. The prior on the augmented $\bm{x}$ therefore includes the adopted prior on $D_{\rm S}$.}
the induced prior $\prior_1(\bm{y})$ is given by
\begin{align}
\label{eq:prior_y_from_x}
\pi_1(\bm{y})=
\pi_1\!\left(\bm{x}=f(\bm{y})\right)
\left|
\frac{\partial \bm{x}}{\partial \bm{y}}
\right|_{\bm{x}=f(\bm{y})}.
\end{align}
Figure~\ref{fig:prior_comp} shows this prior $\pi_1(\bm{y})$ incuded from $\prior_1(\bm{x})$ in orange, together with the Galactic prior $\prior_\gal(\bm{y})$ based on the Koshimoto Galactic model \citep{kos21}, shown in blue.
Although the original prior $\prior_1(\bm{x})$ is simple and apparently uninformative in $\bm{x}$, its image in $\bm{y}$, $\prior_1(\bm{y})$, is highly non-uniform and differs substantially from the Galactic model.
This behavior arises from 
the Jacobian of the transformation in Equation~\ref{eq:prior_y_from_x} \citep{bat11}.
In the present case, it preferentially weights configurations with lower lens masses and shorter lens distances.
When higher-order signals are weak, the likelihood provides only limited information along these directions, so the resulting posterior can be strongly influenced by this induced prior.

In summary, the physically implausible higher-order parameters discussed above can arise from a combination of two effects: imperfections in the likelihood function, such as an inadequate noise model, and the bias introduced by the prior induced in the physical parameter space. 
To examine the latter effect, we test it below using simulated data with a fully specified generative model and purely white noise.


\subsubsection{Simulation Test}\label{sec:simu_test}
We construct a simulated microlensing event, hereafter referred to as Event~1, whose light curve is shown in the left panel of Fig.~\ref{fig:simu_lc}.
To emulate realistic survey observations, we generate light curves in both the $I$ and $V$ bands following the observing cadences of the Microlensing Observations in Astrophysics \citep[MOA;][]{bon01,sum03} and the Korea Microlensing Telescope Network \citep[KMTNet;][]{kim16}.
The simulated data consist of observations from four sites (Data~1--4), with photometric uncertainties assigned according to the empirical error model of the MOA survey.
The purpose of this simulation is to isolate the effect of the prior adopted in the light-curve parameter space.
We therefore generate the photometric scatter as independent Gaussian noise using the assigned uncertainties, without introducing additional time-correlated noise or photometric systematics.
The source has an apparent $I$-band magnitude and color of $(I, V-I)_{\mathrm{S}} = (20, 1)$.

We place the lens in the Galactic disk at $D_{\rm L}=3.66~\mathrm{kpc}$ and adopt a source trajectory crossing a resonant caustic. 
Including microlensing parallax in the model improves the fit by $\Delta\chi^2 \simeq 150$ relative to a static model, so this event would conventionally be regarded as having a statistically significant parallax signal. 
In such a case, one would typically transform the posterior samples obtained in the light-curve parameter space directly into physical parameters, without applying further Bayesian post-processing with a Galactic prior.
The full set of input parameters used to generate this simulated event, including both light-curve parameters and the corresponding physical parameters, is summarized in Table~\ref{table:event2_params}.

Using this data set, we performed MCMC sampling for $p_1(\bm{x}\mid\data)$ in Equation~(\ref{eq:post_x}) using a light-curve model that includes both microlensing parallax and LOM, and assuming uninformative priors on the light-curve parameters as summarized in Table~\ref{table:event2_params}. 
We employed the affine-invariant ensemble sampler \texttt{emcee} \citep{emcee}, using 28 walkers evolved for $2.5\times10^{5}$ steps. The convergence of the MCMC chains is assessed using the integrated autocorrelation time.
We define the effective sample size as $N_{\rm eff}=N/\tau_{\rm int}$, where $N$ is the total number of post-burn-in samples and $\tau_{\rm int}$ is the integrated autocorrelation time.
As a conservative estimate, we use the largest $\tau_{\rm int}$ among the sampled parameters.
With this choice, the chain length corresponds to approximately $50\tau_{\rm int}$ per walker, which is comparable to the commonly used empirical criterion for reliable estimation of the autocorrelation time.
The resulting effective sample size is $N_{\rm eff}\simeq1400$. Thus, at least in the light-curve parameter space, the first-step MCMC provides a sufficiently converged sample for the analysis below.

\subsubsection{Simulation Results}

Table~\ref{table:event2_params} shows the 90\% credible intervals of the inferred light-curve parameters obtained from the above MCMC sampling.
Overall, these intervals cover the input values reasonably well, indicating that the inference in the light-curve parameter space $\bm{x}$ is statistically well behaved.

We now apply the transformation described above to the posterior samples drawn from $p_1(\bm{x}\mid\data)$. 
The resulting samples are then distributed according to
\begin{align}
p_1(\bm{y}\mid\data)
\propto
\mathcal{L}\!\left(\bm{x}=f(\bm{y})\right)\cdot \pi_1(\bm{y}),
\end{align}
where $\pi_1(\bm{y})$ is the induced prior defined in Equation~\ref{eq:prior_y_from_x}.
Figure~\ref{fig:prior_comp} shows the transformed posterior $p_1(\bm{y}\mid\data)$ in red, together with the induced prior $\pi_1(\bm{y})$ in orange.
The corresponding 90\% credible intervals are summarized in Table~\ref{table:event2_params}.

Although the inference in the light-curve parameter space appears statistically well behaved, the transformed posterior in $\bm{y}$ still retains a noticeable dependence on the induced prior $\pi_1(\bm{y})$.
It is shifted toward the region favored by this prior, particularly toward lower lens masses and shorter lens distances, as shown in Figure~\ref{fig:prior_comp}.

Thus, even though this event would conventionally be regarded as having a statistically significant parallax signal, Figure~\ref{fig:prior_comp} shows that the resulting ``direct'' inference can still be biased.
Transforming the light-curve posterior into the physical parameter space remains a Bayesian inference based on $\pi_1(\bm{x})$, and the resulting posterior in $\bm{y}$ therefore retains the influence of the induced prior $\pi_1(\bm{y})$.
This effect is expected to become more pronounced when the parallax signal is weaker, because the likelihood then provides less information with which to overcome the induced prior, making the inference more susceptible to the instabilities described above.

In principle, the effect of such a ``wrong'' prior can be corrected after the first-step inference.
As described in Section~\ref{sec:conv_method}, the conventional two-step approach removes the dependence on the first-step prior by reconstructing the likelihood as in Equation~\eqref{eq:likelihood_from_p1}, and then combines this likelihood with the Galactic prior in the physical parameter space.
This procedure, however, relies on the first-step sampling retaining sufficient likelihood information in the regions that are relevant under the physical prior.
If such regions are strongly disfavored by the induced prior, they may be represented by only a small number of first-step samples, or not sampled at all.
In such poorly sampled regions, the reconstructed likelihood must rely on the assumed form of the approximation rather than being directly constrained by the samples.
This can be particularly problematic when higher-order effects are weakly constrained, because the likelihood may exhibit broad degeneracies and strongly non-Gaussian structure that can be difficult to represent accurately even with flexible Gaussian-mixture approximations such as that of \citet{bac24}.
Thus, although the dependence on the first-step prior can be removed formally, likelihood information that is poorly represented by the finite first-step sample cannot necessarily be recovered accurately in the second step.

\subsection{Computational Inefficiency of the Two-Step Approach}
\label{sec:ineff_twostep}

Even if the likelihood can be accurately reconstructed from the first-step inference, the second step of the conventional approach can itself become computationally inefficient in a high-dimensional parameter space.

As described in Equation~\eqref{eq:post_y}, the second-step inference proceeds by drawing physical parameters $\bm{y}$ from the Galactic prior $\prior_\gal(\bm{y})$ and evaluating the likelihood at the corresponding light-curve parameters $\bm{x}=f(\bm{y})$, using the likelihood reconstructed from the first-step inference through Equation~\eqref{eq:likelihood_from_p1}.
Only those draws from $\prior_\gal(\bm{y})$ that map to regions of high likelihood in the light-curve parameter space contribute appreciably to the final posterior.

Even in commonly used implementations, this step involves resampling over parameters such as $(t_{\rm E}, \theta_{\rm E}, \pi_{\rm E,N}, \pi_{\rm E,E})$, which already defines a moderately high-dimensional space.
If the observational constraints occupy only a small fraction of this space, only a small fraction of samples drawn from $\prior_\gal(\bm{y})$ will map into the relevant high-likelihood region.
As a result, most draws receive negligible importance weights, so the effective sample size becomes very small.
This problem becomes increasingly severe as the constraints become tighter or as the dimensionality of the constrained parameter space increases.

This problem becomes even more severe once LOM is taken into account.
As shown in Appendix~\ref{sec:parameterization}, the LOM parameters are not only related to the orbital elements of the lens but also directly coupled to $D_{\rm L}$ and $D_{\rm S}$.
Therefore, incorporating LOM requires treating the LOM parameters together with other microlensing parameters, which leads to a rapid increase in the effective dimensionality of the resampling problem.
In such a high-dimensional setting, the resampling-based approach becomes extremely inefficient and practically unworkable.

\section{Our Bayesian Framework}\label{sec:method}
We have shown that the use of weakly informative priors in the light-curve parameter space can contribute to the instability and inefficiency of the conventional two-step analysis.
A natural way to avoid this issue is to incorporate a physically motivated Galactic prior from the outset, rather than attempting to correct for it a posteriori.

In principle, this could be achieved by directly sampling the posterior distribution of the physical parameters $\bm{y}$ under the Galactic prior, $p_\gal(\bm{y}\mid\data)$:
\begin{align}
    p_\gal(\bm{y}\mid\data) \propto \mathcal{L}(\bm{y}) \cdot \prior_\gal(\bm{y})
\end{align}
In practice, however, direct sampling in the $\bm{y}$-space is inefficient, because several physical parameters are only weakly constrained by the light curve and exhibit strong degeneracies.
We therefore perform posterior sampling in the $\bm{x}$-space that is more directly constrained by the data, while retaining the Galactic prior defined in the physical parameter space. This leads us to consider the following posterior distribution in $\bm{x}$:
\begin{align}
    p_\gal(\bm{x}\mid\data) \propto \mathcal{L}(\bm{x}) \cdot
    \prior_\gal\!\left(\bm{y}=f^{-1}(\bm{x})\right)
    \left| \frac{\partial \bm{y}}{\partial \bm{x}} \right|_{\bm{y}=f^{-1}(\bm{x})}.
    \label{eq:full_pos}
\end{align}
Here $\left| \partial \bm{y} / \partial \bm{x} \right|$ denotes the absolute value of the Jacobian determinant of the transformation $\bm{y}=f^{-1}(\bm{x})$, and 
ensures proper transformation of the Galactic prior from the physical parameter space $\bm{y}$ to the light-curve parameter space $\bm{x}$.

To realize this approach, two technical challenges must be addressed.
First, we must compute the Jacobian determinant $\left| \partial \bm{y} / \partial \bm{x} \right|$, 
which can be analytically complex due to the nonlinear, coupled transformation between $\bm{x}$ and $\bm{y}$.
Second, we need to evaluate the probability {\it density} $\prior_\gal (\bm{y})$ for arbitrary $\bm{y}$.
This is difficult to obtain efficiently for modern Galactic models based on population synthesis such as \texttt{genulens} \citep{kos21,kos22} and SYNTHPOP\footnote{\url{https://github.com/synthpop-galaxy/synthpop}} \citep{klu25},
because these models typically provide only samples via forward Monte Carlo simulations, not closed-form density functions. 

We developed a tool that addresses both issues. 
Below, we first describe the parameter transformation between the light-curve and physical parameter sets, followed by an efficient approach for computing the Jacobian determinant and a framework that enables fast estimation of the Galactic prior $\prior_\gal(\bm{y})$, referred to as the Galactic Prior Modeling Engine (\texttt{gapmoe}).

\subsection{Parameter Transformation}\label{subsec:params}
In our implementation of the Bayesian framework described above, we work with the following set of binary-lens light-curve parameters,
\begin{align}
\bm{x} = (t_0, u_0, q, \alpha, t_{\rm E}, \rho, \bm{\pi}_{\rm E}, s, \gamma_1, \gamma_2, \gamma_3, r_s, a_s),
\label{eq:full_x}
\end{align}
and the corresponding set of physical parameters,
\begin{align}
\bm{y} = (t_0, u_0, q, M_{\rm L}, D_{\rm L}, D_{\rm S}, \bm{\mu}_{\rm rel}, a, e, i, \Omega_{\rm NE}, \omega, \nu).
\label{eq:full_y}
\end{align}

The parameter set $\bm{x}$ includes the standard microlensing parameters $(t_0, t_{\rm E}, u_0, q, s, \alpha)$ \citep{gau12}, as well as parameters describing higher-order effects: the finite-source effect $\rho$ \citep{wit94,yoo04}, the microlensing parallax vector $\bm{\pi}_{\rm E}$ \citep{gou92,gou00,gou04}, and the LOM parameters $(\bm{\gamma}, r_s, a_s)$ \citep{sko11,boz21}.
The corresponding physical parameter set $\bm{y}$ contains the lens mass $M_{\rm L}$, lens and source distances $(D_{\rm L}, D_{\rm S})$, the relative proper motion vector $\bm{\mu}_{\rm rel}$, and the Keplerian orbital elements $(a, e, i, \Omega_{\rm NE}, \omega, \nu)$ of the lens system, where $a$ is the semi-major axis, $e$ the eccentricity, $i$ the inclination, $\Omega_{\rm NE}$ the longitude of the ascending node measured on the sky from celestial north toward east, $\omega$ the argument of periastron, and $\nu$ the true anomaly.

For this choice of parameterization, the transformation between $\bm{x}$ and $\bm{y}$ is one-to-one. 
In particular, for any given $\bm{x}$, the corresponding physical parameters $\bm{y}=f^{-1}(\bm{x})$ are uniquely determined. 
This one-to-one correspondence is required to evaluate the posterior distribution in Equation~(\ref{eq:full_pos}), which involves transforming the Galactic prior defined in the physical parameter space to the light-curve parameter space via the Jacobian determinant.
The explicit form of the transformation $f^{-1}$ and its derivation are presented in Appendix~\ref{sec:parameterization}.

\subsection{Jacobian Computation using \texttt{JAX}}
As described in Appendix~\ref{sec:parameterization}, the explicit relationship between the parameter sets $\bm{x}$ and $\bm{y}$ can be written in closed form.
Given this relationship, modern automatic differentiation libraries such as \texttt{JAX} \citep{jax18} make it feasible to evaluate the Jacobian $\left| \partial \bm{y} / \partial \bm{x} \right|$ efficiently and accurately.
In this work, we implemented the Jacobian computation using \texttt{jax.jacfwd}, which automatically constructs the Jacobian matrix $\partial \bm{y} / \partial \bm{x}$ at each MCMC step.
The determinant is then evaluated using standard linear algebra routines.
This enables us to incorporate the full Jacobian correction without requiring manual differentiation.

\subsection{Galactic Prior Modeling Engine (\texttt{gapmoe})}\label{sec:gapmoe}

To enable efficient and flexible evaluation of the Galactic prior within our Bayesian microlensing analysis,  
we developed a dedicated computational tool named \texttt{gapmoe}.  
Our implementation is based on the Koshimoto Galactic model \citep{kos21} and its public implementation, \texttt{genulens} \citep{kos22}, which provides a Monte Carlo representation of the Galaxy by sampling lens and source populations according to their spatial distributions, kinematics, and stellar evolution.
Among the full parameter set $\bm{y}$, the Galactic model constrains only the subset $\bm{y}_{\rm Gal}\equiv(M_{\rm L}, D_{\rm L}, D_{\rm S}, \bm{\mu}_{\rm rel})$.

The following describes the procedure by which \texttt{genulens} generates samples of $\bm{y}_{\rm Gal}$.
\begin{enumerate}
    \item \textbf{Source distance:}  
    A source distance $D_{\rm S}$ is drawn from the analytic Galactic stellar density model.  
    This defines the source density profile $\nu_S(D_{\rm S})$ representing the number density of sources along the line of sight.

    \item \textbf{Lens distance:}  
    Given $D_{\rm S}$, a lens distance $D_{\rm L}$ is sampled under the constraint $D_{\rm L} < D_{\rm S}$.  
    The resulting conditional distribution $\nu_L(D_{\rm L}\mid D_{\rm S})$ represents the normalized lens density truncated at $D_{\rm S}$.  

    \item \textbf{Lens mass:}  
    The lens mass $M_{\rm L}$ is drawn from the evolved stellar mass function 
    $f_{M}(M_{\rm L} \mid \tau)$, which reflects the initial mass function (IMF) and 
    stellar evolution processes (e.g., mass loss and remnant formation), and 
    therefore depends on the stellar age $\tau$. 
    The age distribution $p_i(\tau)$ is determined by the Galactic component $i$ (e.g., thin disk, thick disk, or bulge), whose relative contribution at a given lens distance $D_{\rm L}$ is described by the component fraction
    $w_i(D_{\rm L})$.
    Consequently, the overall lens mass distribution depends on $D_{\rm L}$ and given by
    \begin{align}
    g(M_{\rm L} \mid D_{\rm L})
    = \sum_i w_i(D_{\rm L}) f_{M,i}(M_{\rm L}),\label{eq:g_ML}
    \end{align}
    where 
    \begin{align}
    f_{M,i}(M_{\rm L})\equiv\int f_{M}(M_{\rm L} \mid \tau)\, p_i(\tau)\, d\tau,\label{eq:f_ML_i}
    \end{align}
    represents the evolved stellar mass function for each Galactic component $i$. 

    In \texttt{genulens}, a Galactic component $i$ is first selected according to 
    the component fraction $w_i(D_{\rm L})$. 
    Then, a stellar age $\tau$ is drawn from the corresponding age distribution $p_i(\tau)$. 
    An initial mass is sampled from the IMF and evolved according to 
    the stellar age to obtain the present lens mass $M_{\rm L}$. 
    In this way, \texttt{genulens} implements sampling from the lens mass distribution $g(M_{\rm L} \mid D_{\rm L})$.

    \item \textbf{Relative proper motion:}
    The lens and source velocities are sampled independently from their component-specific velocity ellipsoids.  
    The lens--source relative motion $(\mu_{\rm rel}, \varphi_\mu)$ is then computed by subtracting and projecting onto the sky plane, resulting in the distribution $f(\mu_{\rm rel}, \varphi_\mu\mid D_{\rm L}, D_{\rm S})$.

    \item \textbf{Event rate weighting:}
    Each realization is weighted by the microlensing event rate factor
    $D_{\rm L}^2\theta_{\rm E}(M_{\rm L}, D_{\rm L}, D_{\rm S})\mu_{\rm rel}$,
    which accounts for the geometric lensing cross section and the relative lens---source motion (see Section~2 in \citealt{nun25}).
\end{enumerate}
Combining all these terms, the full probability density of the Koshimoto Galactic model can be written as
\begin{align}
p_{\rm kos}(M_{\rm L}, D_{\rm L}, D_{\rm S}, \bm{\mu}_{\rm rel}) 
 &\propto g(M_{\rm L} \mid D_{\rm L}) \cdot \nu_L(D_{\rm L}\mid D_{\rm S}) \cdot \nu_S(D_{\rm S}) \cdot f(\mu_{\rm rel},\varphi_{\mu} \mid D_{\rm L}, D_{\rm S})
 \cdot \frac{1}{\mu_{\rm rel}} \cdot D_{\rm L}^2\,\theta_{\rm E}(M_{\rm L}, D_{\rm L}, D_{\rm S})\,\mu_{\rm rel}.
\end{align}
Here, $1/\mu_{\rm rel}$ is Jacobian factor for the polar-to-Cartesian transformation of $\bm{\mu}_{\rm rel}$.

Based on this factorized structure, \texttt{gapmoe} provides a fast and flexible way to evaluate the Galactic prior $p_{\rm Gal}(\bm{y}_{\rm Gal})$ during Bayesian inference.
We next describe how \texttt{gapmoe} evaluates each term.\\

\noindent\textbf{Source and Lens Distance in \texttt{gapmoe}}

To begin, \texttt{gapmoe} computes the source and lens distance distributions, $\nu_S(D_{\rm S})$ and $\nu_L(D_{\rm L})$, along the line of sight specified by the Galactic coordinates $(l, b)$. 
These are derived from the analytic density models of the Koshimoto Galactic model and are evaluated over the range $[0, 16{,}000]$ pc with a resolution of 1 pc.  
At this stage, $\nu_L(D_{\rm L})$ is computed without enforcing the $D_{\rm L} < D_{\rm S}$ condition.

The discretization of both $\nu_S$ and $\nu_L$ offers several advantages: it reduces computational cost, simplifies normalization, and is consistent with the native 1-pc resolution of the original \texttt{genulens} simulation. 

The conditional distribution $\nu_L(D_{\rm L} \mid D_{\rm S})$ is then obtained by truncating the precomputed $\nu_L(D_{\rm L})$ at $D_{\rm L} < D_{\rm S}$ and renormalizing over this domain.  
Note that $\nu_L(D_{\rm L})$ is marginalized over all Galactic structural components. 
However, for each $D_{\rm L}$, \texttt{gapmoe} also records the fractional contribution of each component $w_i(D_{\rm L})$, which will later be used to construct the lens mass distribution.\\

\noindent\textbf{Lens Mass in \texttt{gapmoe}}

\texttt{gapmoe} computes $f_{M,i}(M_{\rm L})$ 
in Equation~(\ref{eq:f_ML_i}) for each Galactic component $i$ 
using precomputed histograms. 
Each histogram covers the mass range from $M_{\rm L}=0$ to $15\,M_\odot$ 
with a bin width of $\Delta M_{\rm L}=0.01\,M_\odot$. 
They are constructed by performing $10^8$ population-synthesis simulations 
in total, counting the resulting stellar masses into the corresponding bins 
for each Galactic component. 
The precomputed histograms are then weighted by $w_i(D_{\rm L})$ 
to evaluate the probability density $g(M_{\rm L} \mid D_{\rm L})$. 
\\

\noindent\textbf{Relative Proper Motion in \texttt{gapmoe}}

While the stellar velocity distributions of each Galactic component are described analytically in the Koshimoto model, evaluating the joint distribution of the lens-source relative proper motion vector $(\mu_{\rm rel}, \varphi_\mu)$ conditional on $(D_{\rm L}, D_{\rm S})$ is nontrivial.  
This is because $\bm{\mu}_{\rm rel}$ arises from the projected difference between independently sampled lens and source velocities, each drawn from component-specific velocity ellipsoids that depend on position.

To make this tractable, \texttt{gapmoe} approximates the conditional distribution $f(\mu_{\rm rel}, \varphi_\mu \mid D_{\rm L}, D_{\rm S})$ by factorizing it into two independent one-dimensional histograms:
\begin{align}
f(\mu_{\rm rel}, \varphi_\mu \mid D_{\rm L}, D_{\rm S}) \approx f(\mu_{\rm rel} \mid D_{\rm L}, D_{\rm S}) \cdot f(\varphi_\mu \mid D_{\rm L}, D_{\rm S}).
\end{align}
Each of these histograms is defined on a fixed grid:  
$\mu_{\rm rel}$ is binned from $0$ to $22~{\rm mas~yr^{-1}}$ in intervals of $0.5$,  
while $\varphi_\mu$ is binned from $-\pi$ to $\pi$ in $5^\circ$ increments.

These histograms are computed and stored for all combinations of $D_{\rm L}$ and $D_{\rm S}$, 
where $D_{\rm L}$ ranges from $0$ to $15.5~{\rm kpc}$ and $D_{\rm S}$ from $0$ to $16~{\rm kpc}$,
with bin centers placed every $500~{\rm pc}$.  
Only bins satisfying $D_{\rm L} < D_{\rm S}$ are used in the evaluation.  
For each $(D_{\rm L}, D_{\rm S})$ bin, \texttt{gapmoe} performs $10^7$ Monte Carlo simulations to ensure statistical accuracy.  
While this step is computationally intensivee---taking about 8--10 CPU hours 
on a dual-socket 28-core Intel Xeon (56 threads) machine---it only needs to be run once per line of sight and is reused thereafter.

This approximation works well when both the lens and source belong to the bulge, where velocity distributions are nearly isotropic.  
However, in cases where the lens lies in a disk component, anisotropy may induce correlations between magnitude and direction, making the independence assumption less accurate.  
We identify this factorization as a current limitation of \texttt{gapmoe}, and improving the fidelity of the proper motion modeling remains a promising direction for future work.\\
\medskip

\begin{figure}
    \centering
    \includegraphics[width=0.6\textwidth]{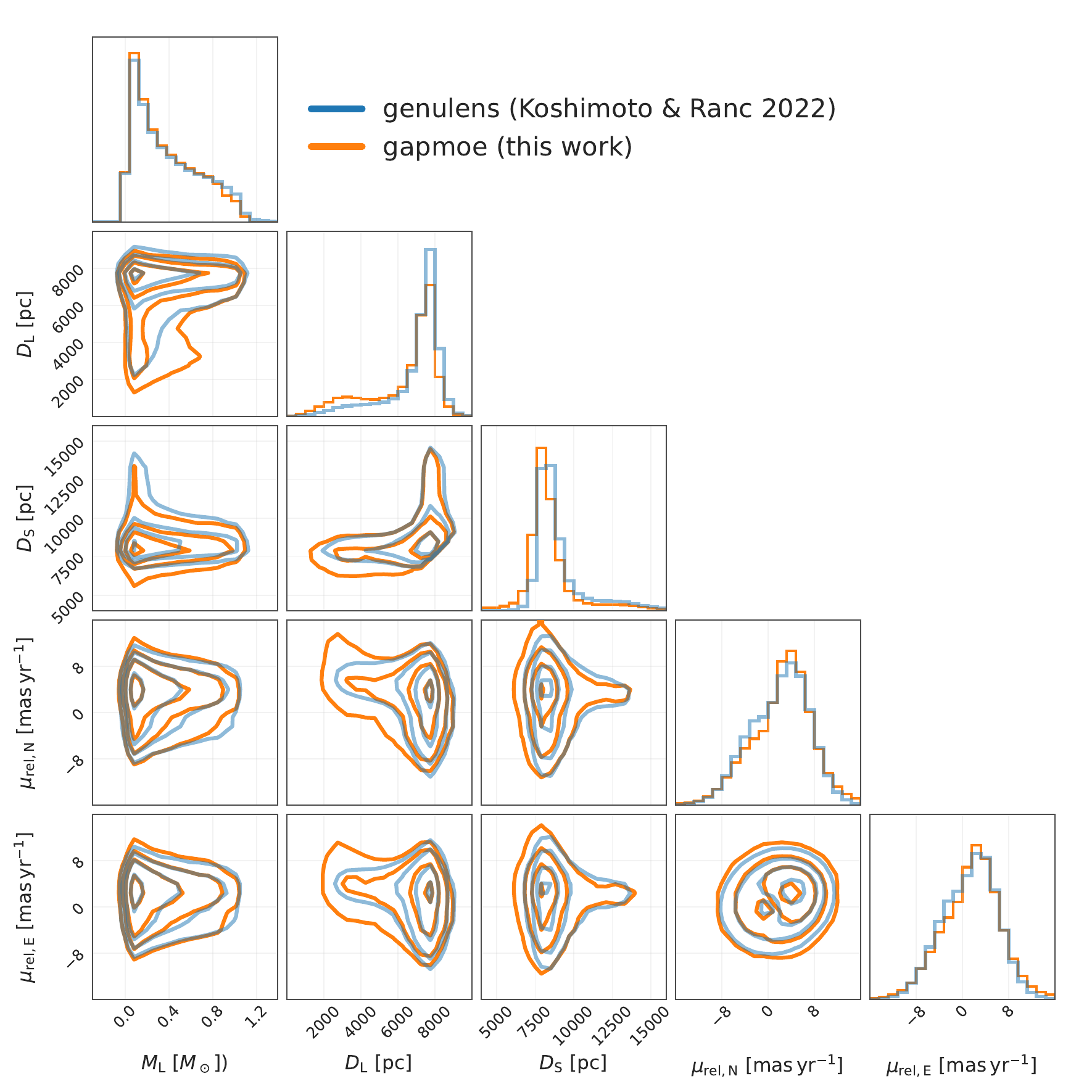}
    \caption{
    Comparison between the direct samples drawn from \texttt{genulens} and those resampled from the density approximation computed by \texttt{gapmoe}.  
    The corner plot shows the marginalized distributions of $M_{\rm L}$, $D_{\rm L}$, $D_{\rm S}$ and $\
    \bm{\mu_{\rm rel}}$.
    }
    \label{fig:comp_gapmoe}
\end{figure}

Combining these analytic evaluation and histogram-based interpolation, the Galactic prior $p_{\rm kos}$ is efficiently approximated within our framework.
To validate the accuracy of our prior evaluation method, we compare in Fig.~\ref{fig:comp_gapmoe} the samples directly generated by \texttt{genulens} with those drawn from the approximate probability density computed by \texttt{gapmoe}.  
Overall, the distributions appear qualitatively similar, confirming that \texttt{gapmoe} provides a reasonable approximation. 

However, notable discrepancies are observed upon closer inspection, particularly in the distribution of $D_{\rm L}$, which shows an excess at small distances ($<5$~kpc) in the \texttt{gapmoe} samples.  
This excess, 
in turn, shifts the $D_{\rm S}$ and $M_{\rm L}$ distributions:  
the $D_{\rm S}$ samples are slightly biased toward smaller values, and the high-mass tail in $M_{\rm L}$ is suppressed,  
consistent with the fact that nearby lenses tend to be lower mass due to their Galactic component.
The origin of this discrepancy lies in the approximation used for the relative proper motion distribution.  
In \texttt{gapmoe}, we assume that the magnitude $\mu_{\rm rel}$ and direction $\varphi_\mu$ of the proper motion vector are independent.  
While this is a good approximation for bulge-bulge lensing (where the distribution is nearly isotropic),  
it fails for disk lenses, where the Galactic rotation introduces directional dependence and leads to larger proper motion magnitudes along preferred directions.  
As a result, the variance of $\mu_{\rm rel}$ is overestimated in the disk-dominated regime,  
which artificially increases the event rate weight $D_{\rm L}^2 \theta_{\rm E} \mu_{\rm rel}$ in these regions and biases the posterior toward closer lenses.
This limitation represents a caveat of the current implementation of \texttt{gapmoe}.

Finally, the numerical tables used by \texttt{gapmoe} are constructed to cover the physically relevant parameter ranges of ordinary Galactic microlensing events. 
The lens-mass histograms are tabulated over $0<M_{\rm L}<15,M_\odot$, the source- and lens-distance distributions extend to $16\,{\rm kpc}$, and the relative-proper-motion magnitude is tabulated over $0<\mu_{\rm rel}<22\,{\rm mas\,yr^{-1}}$. 
For most microlensing events, including the simulated events considered in this work, the relevant parameters are expected to remain within these ranges.
Nevertheless, the edge of a finite numerical table should not be interpreted as a physical hard boundary. 
The lens-mass, distance, and relative-proper-motion-magnitude histograms are therefore extended beyond the boundaries of their positive support using smoothly decaying exponential tails, rather than assigning zero density immediately outside the tabulated range. 
The decay rate at each boundary is estimated from a log-linear fit to up to three positive bins nearest the edge and is bounded to suppress unstable extrapolation caused by Monte Carlo fluctuations.
The distributions are renormalized after including the tails, while physically forbidden regions, such as $M_{\rm L}\leq0$, $D_{\rm L}\leq0$, $D_{\rm S}\leq D_{\rm L}$, and $\mu_{\rm rel}\leq0$, retain zero probability. 
The direction angle $\varphi_\mu$ is treated periodically and is therefore not extrapolated.

\subsection{Comparison with Previous Studies}
Galactic-model information has also been incorporated into several recent microlensing analyses. 
For example, \citet{mro25} constructed and applied an event-specific Galactic-model prior for the microlensing-parallax vector $\bm{\pi}_{\rm E}$, while \citet{rek25} adopted Galactic-model-based priors for the source distance $D_{\rm S}$ and the microlensing parallax. 
In both cases, the information from the Galactic model was projected onto a selected, low-dimensional subset of the event-model parameters. 
In contrast, our framework directly incorporates the correlated joint prior over $(M_{\rm L},D_{\rm L},D_{\rm S},\bm{\mu}_{\rm rel})$ into a general higher-order microlensing inference framework.

Among previous studies, the formulation of \citet{sko11} is conceptually the most similar to ours, in that the Galactic prior is incorporated through an explicit transformation between microlensing observables and physical parameters.
Nevertheless, there are several important differences, which we summarize below.

\subsubsection{Galactic model}
Previous approaches such as \citet{sko11} rely on analytically specified Galactic models, in which the Galactic prior is constructed from simple parametric forms for the stellar density, velocity distribution, and lens mass function (e.g., power-law mass functions such as $\propto M^{-1}$ and Gaussian velocity distributions), allowing the full prior density to be written in closed form.
In contrast, our framework adopts a modern population-synthesis-based Galactic model, specifically the Koshimoto Galactic Model \citep{kos21}, which incorporates stellar evolution, more complex kinematics, and multiple disk components.
Because such models do not admit simple analytic expressions for the full prior density, we employ \texttt{gapmoe} to incorporate these population-synthesis-based Galactic priors in a fully Bayesian manner.

\subsubsection{Using $a_s$ versus using $D_{\rm S}$.}
The calculation of the magnification including LOM requires specifying the semi-major axis normalized by the Einstein radius, $a/R_{\rm E}$. In \citet{sko11}, this quantity is obtained indirectly by treating $D_{\rm S}$ as a fundamental parameter and then using the parameter transformation to determine $a_s$. 
Although this procedure is mathematically possible, it is not straightforward in practice because the magnification cannot be computed from $D_{\rm S}$ alone.

Specifically, recovering $a_s$ from Equation~\eqref{eq:DS} requires the Einstein radius $\theta_{\rm E}$, which depends on the angular source radius $\theta_*$. 
However, $\theta_*$ is inferred from the source flux, and the source flux is itself determined through the magnification model. 
Thus, if $D_{\rm S}$ is taken as a fundamental model parameter, the computation of the magnification involves an implicit circular dependence. 
In \citet{sko11}, this issue is avoided by fixing $\theta_*$ to a fiducial value, rather than allowing it to vary self-consistently with the source flux implied by the magnification model.

In contrast, we adopt the parameterization of \citet{boz21}, in which $a_s$ is introduced as the ratio of the semi-major axis to the projected binary separation at the reference time. 
Since the projected separation is expressed in units of $R_{\rm E}$, $a_s$ directly specifies $a/R_{\rm E}$. 
The magnification can therefore be computed without prior knowledge of $\theta_*$, allowing $\theta_*$ and $D_{\rm S}$ to be inferred consistently from the light curve without any fiducial assumption.

\subsubsection{Definition of the longitude of the ascending node}
In \citet{sko11}, the nodal angle is defined relative to the binary axis at the reference time. 
In the notation of Appendix~\ref{sec:parameterization}, this angle corresponds to $\Omega_0$.
Because this reference axis is itself determined by the orbital configuration, $\Omega_0$ is not an independent degree of freedom specifying the orientation of the orbit in a fixed sky frame.

In this work, we instead define the longitude of the ascending node with respect to a fixed sky frame. 
Specifically, we use $\Omega_{\rm NE}$, measured on the sky from celestial north toward east. 
With this definition, $\Omega_{\rm NE}$ independently specifies the orientation of the line of nodes on the sky. 
Together with $(a,e,i,\omega,\nu)$, it provides a complete specification of the orbital configuration in the fixed sky frame, so that the physical parameter set $\bm{y}$ uniquely determines the corresponding orbit.

This choice avoids the ambiguity that arises when the nodal angle is defined relative to the binary axis rather than to an external reference direction. 
A more detailed discussion, including the implication for the Jacobian, is given in \citet{2026arXiv260512982M}.

\section{Application to Simulated Data} \label{sec:result_new}
We apply our framework to a simulated event where the higher-order effects are weak, and compare the results with traditional two-step Bayesian analysis.

Our goal is twofold.
First, we examine whether the proposed framework can robustly recover the correct physical parameters in a regime where higher-order effects are weak or only marginally constrained by the light curve.
Second, we compare the results with those obtained using the conventional two-step approach, examining how the inferred lens mass and distance change when weak higher-order effects are consistently included, in comparison with the conventional two-step approach.

\subsection{Event description}
As an example of an event with weak higher-order effects, we consider a second simulated event, hereafter referred to as Event~2. 
The simulated light curve is shown in the right panel of Figure~\ref{fig:simu_lc}, and the input parameters are listed in Table~\ref{table:event1_gap_wo}.
The observational setup and the treatment of photometric uncertainties are identical to those described in Section~\ref{sec:simu_test}.
The source has a color and magnitude of $(I, V-I)_\mathrm{S} = (20, 1)$.
The lens is located in the Galactic bulge at $D_{\rm L} = 6.69~\mathrm{kpc}$, which makes the microlensing parallax effect intrinsically weak.
In addition, the caustic-crossing feature occurs on a short timescale of $\sim0.5$~days, limiting the sensitivity to LOM.
As a result, even when both parallax and LOM are included in the model, the improvement in fit relative to the static model remains small ($\Delta\chi^2 \lesssim 2$).
Under the conventional $\chi^2$ threshold-based approach, such weak higher-order signals would therefore be ignored.

\subsection{Inference setup}

The purpose of this section is to examine whether our proposed framework can stably recover the physical parameters even when the higher-order effects are weak, and to compare the results with those of the conventional two-step approach when these effects are treated consistently. 

To this end, we analyze this simulated event using two inference schemes.
In the first scheme, we apply our Bayesian framework and directly sample the full posterior distribution given by Equation~\eqref{eq:full_pos} using \texttt{gapmoe}. Microlensing parallax and lens orbital motion with an eccentric orbit are included from the outset. The prior distributions used in this analysis are those listed in Table~\ref{table:event1_gap_wo}.
In the second scheme, we perform a conventional two-step analysis. First, we fit the light curve with a model that ignores parallax and lens orbital motion, and sample the light-curve parameters $\bm{x}$ from the posterior distribution given by Equation~\eqref{eq:post_x}. The priors adopted in this step are those listed in Table~\ref{table:event1_gap_wo}. Second, we construct the posterior distribution of the physical parameters $\bm{y}$ by combining a Galactic prior with an approximate likelihood based on the posterior samples of $\bm{x}$ obtained in the first step. Following the conventional procedure, this approximate likelihood is modeled as a product of independent Gaussian distributions in $t_{\rm E}$ and $\theta_{\rm E}$.

Posterior sampling in both analyses is performed using \texttt{emcee} \citep{emcee}, following the same setup and convergence diagnostics as in Section~\ref{sec:simu_test}.
We use 28 walkers evolved for $2.5\times10^{5}$ steps.
Using the effective sample size definition based on the integrated autocorrelation time introduced in Section~\ref{sec:simu_test}, we obtain $N_{\rm eff}\simeq2.4\times10^3$ for the \texttt{gapmoe} analysis and $N_{\rm eff}\simeq2.7\times10^3$ for the first-step light-curve MCMC in the conventional two-step analysis.
These values correspond to chain lengths of approximately $90\tau_{\rm int}$ and $100\tau_{\rm int}$ per walker, respectively.
Thus, both analyses provide a sufficient number of effectively independent samples for the comparison below.

\subsection{Test result}
\begin{figure}[t]
    \centering
    \includegraphics[width=0.85\textwidth]{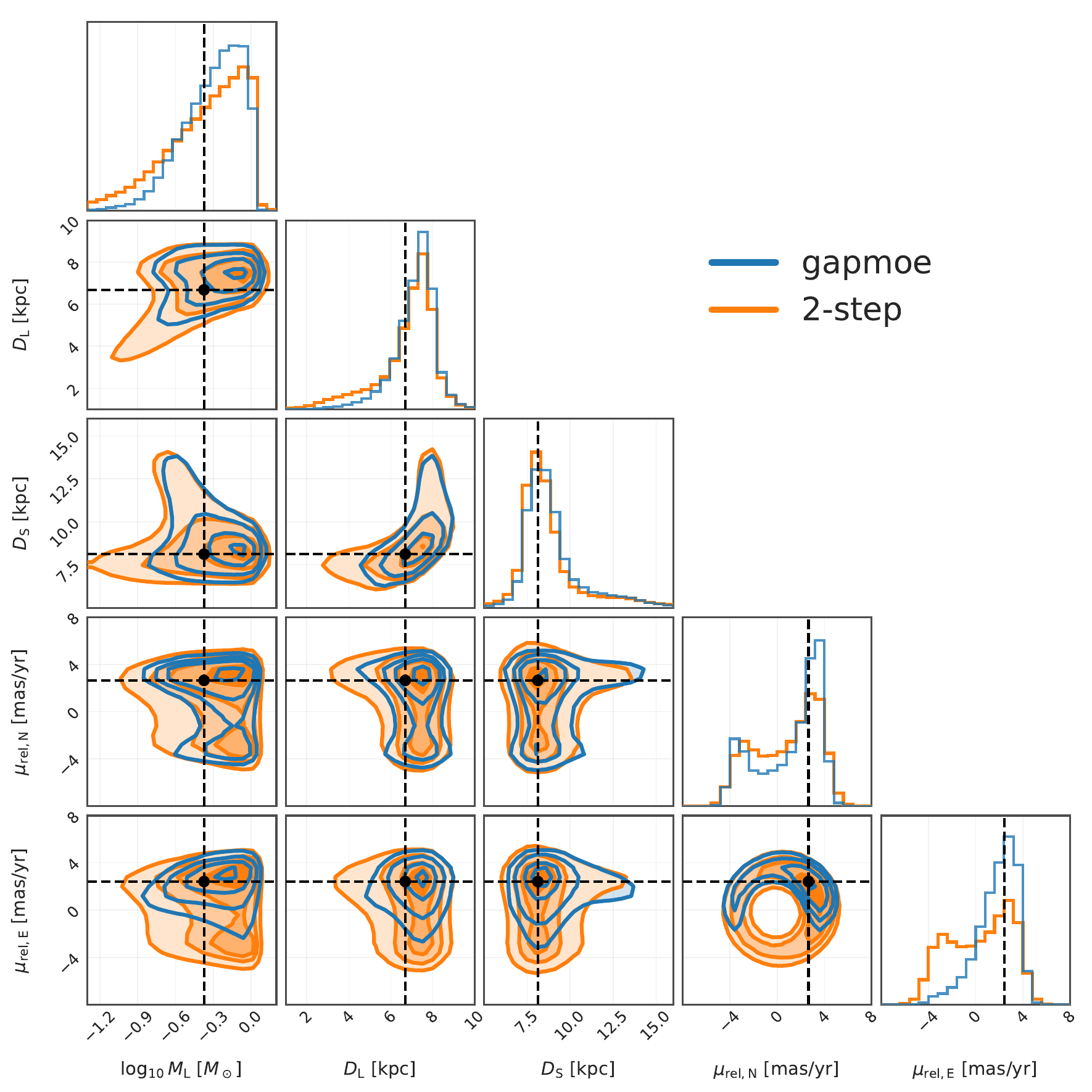}
    \caption{
    Comparison of posterior distributions for selected physical parameters of Event~2.
    The blue contours show the posterior obtained with our \texttt{gapmoe}-based framework, while the orange contours show those from the conventional two-step analysis.
    Black dashed lines indicate the true input values.
    Other parameters are summarized in Table~\ref{table:event1_gap_wo}.
    }
    \label{fig:sim_result_event1}
\end{figure}

Figure~\ref{fig:sim_result_event1} shows the corner plot of the posterior distributions for the physical parameters $(M_{\rm L}, D_{\rm L}, D_{\rm S}, \mu_{\rm rel,N}, \mu_{\rm rel,E})$ inferred from the simulated event.
The black dashed lines indicate the true input values.
Table~\ref{table:event1_gap_wo} summarizes the central 90\% credible intervals for all inferred parameters.

The posterior distributions obtained with \texttt{gapmoe} (blue contour) substantially overlap with those from the conventional two-step analysis (orange contour) and also include the true input values.
This shows that our framework can incorporate higher-order effects directly even in cases where the conventional approach would need to ignore them, while still enabling robust physical inference.

At the same time, the \texttt{gapmoe} posteriors are modestly but systematically tighter. 
This improvement is most evident in the relative proper motion $\bm{\mu}_{\rm rel}$, whose directional information is not constrained in the conventional two-step analysis because it is encoded in the parallax signal that is ignored there. 
As a result, the corresponding posterior remains largely prior dominated in the conventional approach, whereas our framework yields a more informative constraint by consistently incorporating the weak higher-order signals.
More modest improvements are also seen for the lens mass $M_{\rm L}$ and distance $D_{\rm L}$.

These results indicate that the proposed framework successfully extends the conventional inference scheme to the regime of weak higher-order effects: it remains consistent with the conventional results when those effects are weak, while still extracting the additional information they contain.

\begin{deluxetable*}{cc|cc|cc|c}
\tablecaption{Summary statistics for simulated Event~2 with the \texttt{gapmoe} method and the 2-step method: central 90\% intervals (p5--p95), true values, and prior distributions.\label{table:event1_gap_wo}}
\tablehead{
\colhead{Parameter} & \colhead{Unit} &
\multicolumn{2}{c|}{\texttt{gapmoe}} &
\multicolumn{2}{c|}{2-step} &
\colhead{True} \\
&
&
\colhead{90\% interval} & \colhead{Prior} &
\colhead{90\% interval} & \colhead{Prior} &
}
\startdata
\cutinhead{Light-curve parameters}
$t_0 - 10085$ & day & $-0.049\,\text{--}\,0.035$ & $\mathcal{U}(-10,10)$ & $-0.046\,\text{--}\,0.039$ & $\mathcal{U}(-10,10)$ & $0.000$ \\
$t_{\rm E}$ & day & $29.10\,\text{--}\,30.16$ & --- & $29.08\,\text{--}\,30.11$ & $\mathcal{U}(25,35)$ & $30.00$ \\
$u_0$ & $10^{-2}$ & $0.93\,\text{--}\,1.11$ & $\mathcal{U}(-500,500)$ & $0.92\,\text{--}\,1.11$ & $\mathcal{U}(-500,500)$ & $1.00$ \\
$\rho$ & $10^{-3}$ & $0.44\,\text{--}\,0.68$ & --- & $0.46\,\text{--}\,0.71$ & $\mathcal{U}(0,1)$ & $0.60$ \\
$q$ & $\log_{10}$ & $-3.79\,\text{--}\,-2.70$ & $\mathcal{U}(-5,0)$ & $-3.72\,\text{--}\,-2.68$ & $\mathcal{U}(-5,0)$ & $-3.00$ \\
$s$ & --- & $1.074\,\text{--}\,1.184$ & --- & $1.080\,\text{--}\,1.191$ & $\mathcal{U}(0,5)$ & $1.145$ \\
$\alpha$ & rad & $2.488\,\text{--}\,2.636$ & --- & $2.482\,\text{--}\,2.634$ & $\mathcal{U}(-\pi,\pi)$ & $2.571$ \\
$\pi_{\rm E,N}$ & $10^{-1}$ & $-0.88\,\text{--}\,1.66$ & --- & $\text{---}$ & --- & $0.64$ \\
$\pi_{\rm E,E}$ & $10^{-1}$ & $-0.24\,\text{--}\,1.17$ & --- & $\text{---}$ & --- & $0.60$ \\
$\gamma_1$ & $10^{-3}~{\rm day}^{-1}$ & $-2.48\,\text{--}\,2.68$ & --- & $\text{---}$ & --- & $0.32$ \\
$\gamma_2$ & $10^{-3}~{\rm day}^{-1}$ & $-3.38\,\text{--}\,3.25$ & --- & $\text{---}$ & --- & $-0.31$ \\
$\gamma_3$ & $10^{-3}~{\rm day}^{-1}$ & $-3.13\,\text{--}\,3.15$ & --- & $\text{---}$ & --- & $3.19$ \\
$r_s$ & --- & $-2.23\,\text{--}\,2.10$ & --- & $\text{---}$ & --- & $0.00$ \\
$a_s$ & --- & $0.68\,\text{--}\,7.15$ & --- & $\text{---}$ & --- & $1.02$ \\
\cutinhead{Physical parameters}
$M_{\rm L}$ & $M_\odot$ & $0.18\,\text{--}\,0.99$ & Galactic model & $0.09\,\text{--}\,1.02$ & Galactic model & $0.42$ \\
$D_{\rm L}$ & kpc & $5.24\,\text{--}\,8.46$ & Galactic model & $3.70\,\text{--}\,8.38$ & Galactic model & $6.69$ \\
$D_{\rm S}$ & kpc & $7.09\,\text{--}\,12.93$ & Galactic model & $6.73\,\text{--}\,12.97$ & Galactic model & $8.12$ \\
$\mu_{\rm rel,N}$ & mas/yr & $-3.71\,\text{--}\,4.08$ & Galactic model & $-3.69\,\text{--}\,4.33$ & Galactic model & $2.65$ \\
$\mu_{\rm rel,E}$ & mas/yr & $-1.57\,\text{--}\,3.96$ & Galactic model & $-3.86\,\text{--}\,3.98$ & Galactic model & $2.42$ \\
$a$ & AU & $0.25\,\text{--}\,1.47$ & \text{log}~$\mathcal{U}(10^{-1},10^{4})$ & $\text{---}$ & --- & $2.34$ \\
$e$ & --- & $0.05\,\text{--}\,0.93$ & $\mathcal{U}(0,1)$ & $\text{---}$ & --- & $0.10$ \\
$\cos{i}$ & --- & $-0.90\,\text{--}\,0.90$ & $\mathcal{U}(-1,1)$ & $\text{---}$ & --- & $-0.10$ \\
$\omega$ & rad & $-2.83\,\text{--}\,2.83$ & $\mathcal{U}(-\pi,\pi)$ & $\text{---}$ & --- & $-1.81$ \\
$\Omega_{\rm NE}$ & rad & $-2.79\,\text{--}\,2.75$ & $\mathcal{U}(-\pi,\pi)$ & $\text{---}$ & --- & $-1.51$ \\
$\nu$ & rad & $-2.82\,\text{--}\,2.85$ & $\mathcal{U}(-\pi,\pi)$ & $\text{---}$ & --- & $1.51$ \\
\enddata
\end{deluxetable*}

\section{Discussion}
\label{sec:discussion}
\subsection{Higher-Order Effects: To Include or Not To Include?}\label{sec:model_selection}

Many previous investigators adopted a cautious approach to avoid making the model unnecessarily complex unless the data provide `strong' support for the presence of higher-order effects in the light curves. 
While this approach makes sense, its accuracy crucially relies on the success of model selection --- a difficult task when the signals are weak, statistical improvements are only marginal, and light-curve systematics are not fully understood.
As a result, implementing this strategy in a reliable manner remains challenging. A common practice is to use $\Delta\chi^2$ with respect to a model without higher-order effects to decide whether additional parameters are warranted.  
Yet such thresholds are essentially ad-hoc, since $\chi^2$ alone does not provide a sufficient basis for model selection.  
Moreover, photometric errors are rarely perfectly independent and Gaussian, so the true likelihood deviates from the $\exp(-\chi^2/2)$ form as given in Equation~\eqref{eq:likelihood_normal}. 
Consequently, the statistical interpretation of $\Delta\chi^2$ values becomes subtle, necessitating relatively large thresholds to claim a detection of higher-order effects.  

This post-hoc decision-making process is impractical, particularly in the Roman era where a uniform, automated analysis of thousands of events will be required.  
The reliance on arbitrary thresholds also complicates population-level studies, because event-by-event decisions about whether to include higher-order effects introduce a selection that is difficult to characterize consistently across the population (Section~\ref{sec-intro}). 
Our Bayesian framework addresses these issues by enabling inference even when higher-order effects are not strongly apparent in the light curve.  
This strategy is also physically well motivated: parallax and orbital motion are always present in binary-lens events, and it is therefore more natural to account for them consistently, rather than to rely on uncertain model selection criteria.

\subsection{Roles of Priors in Event- and Population-Level Inference}
\label{sec:prior_roles}

Our framework adopts a Galactic prior as a default choice. Our rationale for favoring it over a prior uniform in the microlensing parameter space, $\prior_1$, is that it generally provides a less misleading representation of the physical parameter distribution and often provides a more efficient summary of the data, while preserving the information contained in the likelihood. 
Here we revisit these points from slightly different viewpoints by asking whether adopting a Galactic prior may introduce additional model dependence or bias into the inference. The discussions are also intended to clarify the roles of the prior in individual and population-level analyses.

\subsubsection{Is an ``Uninformative'' Prior Safer?}\label{ssec:uninformative_prior}

Galactic models are constructed using observational constraints together with assumptions about, for example, the spatial distribution, kinematics, and mass distribution of Galactic stellar populations.
Consequently, any inaccuracies in the assumptions made in constructing the model, limitations of the adopted parameterization, and systematic uncertainties in the data used to constrain it could call into question the reliability of a Bayesian inference result when the data are weakly constraining.

This is a legitimate concern. It is therefore useful to follow a general practice in Bayesian inference and examine the sensitivity of the results to the adopted prior. For example, comparisons using different Galactic models or different assumptions within a given model can help distinguish constraints that are robustly provided by the light curve from those that depend more strongly on the adopted population model.
This is especially important when interpreting the inferred properties of individual systems.

That said, we also note that an ``uninformative'' prior does not necessarily resolve this concern. It can instead be a more dangerous choice: as we pointed out in Section~\ref{sec:noninformative_prior_bias}, a prior uniform in the microlensing parameters $\bm{x}$ is strongly informative in the space of physical parameters $\bm{y}$, where the induced prior is most likely ``more wrong'' than Galactic priors.
A safer alternative would be to adopt a prior that is uniform directly in the physical parameters $\bm{y}$ (or the prior for $\bm{x}$ induced from that, in practice),  thereby avoiding this particular coordinate-induced distortion.
Such a choice may indeed be useful if one wishes to minimize dependence on a specific Galactic model. 

More generally, however, there is no guarantee that weakly informative priors provide less biased results when bias is considered over an ensemble of events drawn from the underlying population.
For example, when the underlying population of physical parameters is strongly non-uniform, a combination of a uniform prior and a weakly constraining likelihood function results in posterior estimates that tend to assign more probability to regions of the parameter space that are less representative of the underlying population.
In the context of our example, the stellar parameters are generally drawn from a joint distribution that is by no means uniform, and so are the microlensing parameters converted from them. Therefore, an inference based on a Galactic prior tends to be less misleading than that based on a weakly informative one, at least in a population-level sense.\footnote{What if there is an individual event that lies far in the tail of the population described by Galactic models? Should we be concerned that an inference based on a Galactic prior might unfortunately miss such an interesting event? 
We do expect that the inference based on a Galactic prior generally assigns less posterior probability to such a solution than an analysis based on an ``uninformative'' prior.
But it makes sense to do so: if an apparently rare and interesting event is identified, one should weigh the evidence for such a claim contained in the data against our prior expectation to see if it's strong enough to justify that claim. A Bayesian analysis with a Galactic prior provides a quantitative, self-consistent means to do so. It's still true, though, that the prior needs to be designed carefully so that such solutions are not completely excluded a priori by assigning exactly zero prior probability density. Our prescription for the prior edges described at the end of Section~\ref{sec:gapmoe} is designed to better handle such edge cases.}

\subsubsection{Posterior Samples as a Summary of the Data}\label{ssec:posterior_as_a_summary}

One might instead argue that $p_1(\bm{x}|D)$ conditioned on an uninformative prior $\pi_1$ --- which appears to involve fewer assumptions than a Galactic prior --- represents a more ``faithful'' or ``neutral'' summary of the data. We argue that this intuition is not generally justified.

The Bayesian posterior is a probability density over the model parameters, and therefore transforms with a Jacobian under a change of parameterization, unlike the likelihood. It is thus always a prior-dependent summary of the data, even for a uniform prior. As we have demonstrated, whether a given prior is uniform depends explicitly on the parameterization. Even when one adopts a uniform prior (sometimes informally described as ``no prior'') in a particular parameterization, the same prior is generally non-uniform under another parameterization. As is widely recognized, this choice matters little in practice when the data are highly constraining, but it can matter substantially in a weakly constrained regime.

This means that both the posterior and prior need to be specified to fully represent the information encoded in the given data, i.e., the likelihood function. Conversely, if both are known, the likelihood function can in principle be recovered, up to an overall normalization, by dividing the posterior density by the prior density (i.e., Equation~\ref{eq:likelihood_from_p1}), given adequate posterior sampling and prior support over the relevant region of parameter space.

Thus, in the idealized limit, different prior choices all lead to a valid summary of the data, and no prior is intrinsically more faithful to the data than another.
In practice, however, we can draw only a finite number of samples, and different choices lead to different efficiencies: some priors lead to more posterior samples being drawn from certain regions of parameter space than from others, and the posterior samples will provide a more efficient summary when more are drawn from the most relevant regions of parameter space. The issue we pointed out in Section~\ref{sec:lim_conv_method} is that the summary can become too inefficient to be useful when the prior puts most of the probability mass in the irrelevant region. In this sense, a Galactic prior is no more faithful than an uninformative one, but it can be more efficient in summarizing the information in the data.

This raises the next question: what prior is the most efficient? In other words, what defines the ``relevant'' region of the parameter space that should be resolved most finely? The answer depends on the purpose for which the posterior samples will be used.
One important use --- which we discuss further in the next subsection --- is to infer population-level properties from a collection of microlensing events, such as the distribution of lens masses along a given line of sight \citep[e.g.,][]{sum23} or the occurrence rate of planets as a function of host and planet properties \citep[e.g.,][]{suz16,zan25}. As we make explicit below,
the most useful samples in this case are those that resolve well the regions of parameter space that are frequently populated by the actual ensemble of events. However, this underlying population is itself what we ultimately wish to learn from the data and is therefore not known a priori.
In this sense, the statement above should be interpreted with some caution. A more precise statement is that a Galactic prior likely provides a reasonable initial guess for where those regions are likely to lie, and would therefore be more efficient than the highly distorted prior $\pi_1(\bm{y})$ induced by a uniform prior in $\bm{x}$.
It may even be more efficient than a uniform prior in $\bm{y}$, if the Galactic prior provides a reasonably good approximation to the underlying population distribution.
We make this point more explicit in the next subsection using a hierarchical formulation.

\subsubsection{Population-level Inference}\label{sec:population_inference}

The interpretation of weakly constrained individual events ultimately depends on the prior, and a decisive conclusion may not be possible from an individual event alone. However, the information can still be useful for inferring the population-level properties of the events, such as the distribution of lens properties represented by Galactic models, the distribution of planet properties, or both.

One convenient formulation of such a problem is to adopt a hierarchical model \citep[e.g.,][]{2010ApJ...725.2166H,2014ApJ...795...64F}. Although constructing a specific and practical framework is beyond the scope of this work, we give a rough sketch of the formulation to make what we meant above more explicit.
For simplicity, in this schematic discussion we suppress survey selection effects, including detection efficiencies, as well as details of how the parameters are defined for events with and without planets.\footnote{
In a complete analysis, the precise definition of $\data_i$ and the corresponding likelihood depend on whether one works directly with the light curves or with catalog-level detection/non-detection information. In the latter case, for example, detections, non-detections, and detection efficiencies could be incorporated through a Poisson point-process likelihood \citep[e.g.,][]{1997ApJ...486..697A,miy23}. We suppress these details here because they do not affect the schematic hierarchical structure discussed below.
}
Suppose we wish to infer the population-level distribution $p(\bm{y}\mid\boldsymbol{\eta})$ from a collection of microlensing events labeled by $i$, which can include events both with and without planets. Here $\bm{y}$ may include parameters such as those in Equation~\ref{eq:full_y} (when the lens has a planet), and examples of $\bm{\eta}$ include the parameters of the planet occurrence function and the parameters describing the lens parameter distribution within the Galaxy.
Then, schematically, the joint posterior can be written as
\begin{align}
    \label{eq:hbayes_full} p(\{\bm{y}_i\},\boldsymbol{\eta}\mid\{\data_i\}) \propto p(\boldsymbol{\eta}) \prod_i \mathcal{L}_i\left(f(\bm{y}_i)\right) p(\bm{y}_i\mid\boldsymbol{\eta})
\end{align}
when given the data $\data_i$ for each event.
This represents the full joint inference at this schematic level, but it involves all event-level and population-level parameters simultaneously.
One practical path --- which also corresponds to the idea of using posterior samples as summaries of the information in the data discussed above ---
is to focus on the population-level parameters $\boldsymbol{\eta}$ alone by marginalizing over the individual $\bm{y}_i$ to obtain
\begin{align}
    \label{eq:hbayes_marg}
p(\boldsymbol{\eta}\mid\{\data_i\})
\propto
p(\boldsymbol{\eta})
\prod_i
\int
\mathcal{L}_i\!\left(f(\bm{y}_i)\right)
p(\bm{y}_i\mid\boldsymbol{\eta})
\,\mathrm{d}\bm{y}_i.
\end{align}
Then the last integral can be evaluated by recycling the posterior samples from the individual-event analyses \citep{2010ApJ...725.2166H}: if the posterior samples for event $i$ were obtained using a certain prior $\pi_i(\bm{y}_i)$, so that $p_i(\bm{y}_i\mid\data_i) \propto \like_i(f(\bm{y}_i))\,\prior_i(\bm{y}_i)$, then
\begin{align}
    \label{eq:hbayes_IS} 
    \int \like_i \left(f(\bm{y}_i)\right) p(\bm{y}_i\mid\boldsymbol{\eta}) \,\mathrm{d}\bm{y}_i \propto \int p_i(\bm{y}_i\mid\data_i) \frac{p(\bm{y}_i\mid\boldsymbol{\eta})}{\pi_i(\bm{y}_i)} \,\mathrm{d}\bm{y}_i \approx {1\over K_i} \sum_{k=1}^{K_i} \frac{p(\bm{y}_i^{(k)}\mid\boldsymbol{\eta})}{\pi_i(\bm{y}_i^{(k)})}, 
\end{align}
where $\bm{y}_i^{(k)}$ are $K_i$ samples from the individual-event posterior conditioned on the prior $\prior_i$.
In the present framework, the samples obtained from Equation~\eqref{eq:full_pos}, after transformation to $\bm{y}$, provide these individual-event posterior samples, allowing the higher-order information retained in the individual-event analyses to be propagated directly into a population-level inference without reducing each event to a small set of summary quantities.
Examples of such hierarchical analyses can be found in related contexts.
These include inference of the mass function of dark lenses \citep{2020A&A...636A..20W,2024ApJ...961..179P}, as well as inference of planet occurrence rates as functions of planetary parameters \citep{2014ApJ...795...64F,pol21}, stellar or host parameters \citep{nun24}, and both sets of parameters \citep{miy23}, while accounting for uncertainties in the relevant event- or object-level parameters.\footnote{In the microlensing context, the Galactic model, or more generally the population distribution of lens parameters, is usually held fixed in such analyses. In principle, however, one can instead infer the Galactic population distribution together with the planet population by jointly analyzing events with and without planets. All events then constrain the lens-property distribution; planetary and non-planetary events together constrain the occurrence rate, and the planetary events additionally constrain the distribution of planet properties. This does not double-count the information or introduce a circularity: the Galactic and planetary population parameters are simply updated jointly within a single generative model.
}

An important feature of Equations~\ref{eq:hbayes_full} and \ref{eq:hbayes_marg} is that the information for each event enters as a likelihood function, which is independent of the prior used for posterior sampling of individual events. This remains the case even in Equation~\ref{eq:hbayes_IS}, which apparently includes posterior samples:
the individual-event posterior enters through the ratio $p_i(\bm{y}_i|\data_i) / \pi_i(\bm{y}_i) \propto \mathcal{L}_i(f(\bm{y}_i))$, so the explicit dependence on the prior used in the individual-event analysis is removed.
The prior in the individual-event analysis therefore serves only as an ``interim'' distribution for summarizing the information in the data.
At the same time, for a fixed number of posterior samples $K_i$, the computation is most efficient when the importance weights $p(\bm{y}_i|\bm{\eta})/\pi_i(\bm{y}_i)$ in the integrand of Equation~\ref{eq:hbayes_IS} vary little across the samples: when they vary strongly, only a small fraction of the samples effectively contribute to the integral, reducing the effective sample size.
Thus, if one wishes to recycle the individual posterior samples for this type of analysis, the choice of the interim prior is formally a matter of computational efficiency, provided that it adequately covers the region required by the population model \citep[cf.][]{2010ApJ...725.2166H}.
Thus, an efficient choice of interim prior $\prior_i(\bm{y})$ would be one similar to the target population distribution $p(\bm{y}|\boldsymbol{\eta})$. We argue that a Galactic model provides a reasonable initial guess for such a distribution.
This provides a formal formulation of the argument made in Section~\ref{ssec:posterior_as_a_summary}.

\subsection{Role of \texttt{gapmoe} in Two-step Inference}

As discussed in Section~\ref{sec:lim_conv_method}, the conventional two-step approach can become problematic when higher-order effects are weak.
Our main framework avoids this issue by incorporating the Galactic prior directly into the light-curve inference.

The usefulness of \texttt{gapmoe}, however, is not limited to this direct-inference framework.
As discussed in Section~\ref{sec:ineff_twostep}, the second-stage inference can become highly inefficient when only a small fraction of samples drawn from the Galactic prior are consistent with the observational constraints.
Because \texttt{gapmoe} allows the Galactic prior density to be evaluated directly at arbitrary physical parameters, the second-stage inference can instead be formulated as an MCMC problem, so that the sampler can directly explore regions of high posterior probability without relying on rejection sampling.

Therefore, \texttt{gapmoe} is useful not only for incorporating a Galactic prior directly into light-curve inference, but also as a density model that improves Galactic-prior-based inference within a two-step framework.
This advantage may be particularly important for catalog-level statistical analyses based on large surveys.

\section{Summary and Future Prospects}
\label{sec:summary}
\subsection{Summary}
We clarified the limitations of the conventional two-step Bayesian analysis of binary-lens events. 
When higher-order effects such as microlensing parallax and lens orbital motion are included, an apparently uninformative prior in the light-curve parameter space can introduce significant biases in the inferred physical parameters, and this effect can remain non-negligible even when the higher-order signals are moderately constrained (Section~\ref{sec:lim_conv_method}).
The same prior can also drive MCMC sampling toward physically implausible regions while undersampling the physically relevant ones favored by a Galactic prior.
As a result, posterior reweighting becomes highly inefficient and may fail to recover the physical posterior accurately when the relevant regions are poorly sampled.

To address this, we introduced a physically consistent Bayesian framework that directly incorporates a Galactic prior during sampling in light-curve parameters such as $t_{\rm E}$.
This was made possible by two technical developments: (1) accurate computation of the full Jacobian $\left| \partial \bm{y} / \partial \bm{x} \right|$ using automatic differentiation, and (2) an efficient method to evaluate the physically motivated prior for $(M_{\rm L}, D_{\rm L}, D_{\rm S}, \bm{\mu}_{\rm rel})$ based on forward simulations using our new tool \texttt{gapmoe}.

We demonstrated this approach on a synthetic microlensing event with weak higher-order effects and found that our method robustly recovered the correct physical solution even when the higher-order signals were only marginally constrained by the data.
This robustness eliminates the need for per-event model selection, allowing for a unified modeling approach that consistently includes higher-order effects even when they are only weakly constrained.

\subsection{Future Applications and Prospects}
\label{sec:future_prospects}

Our proposed framework addresses key shortcomings of conventional two-step microlensing analysis and provides a unified approach to inference that is both physically motivated and scalable.
This makes it particularly valuable in the context of upcoming large-scale microlensing surveys such as the \textit{Roman Space Telescope}, which will detect and monitor thousands of microlensing events.

From a computational perspective, a fully differentiable representation of the Galactic prior density presented in this paper would enable seamless integration with automatically differentiable microlensing light-curve models \citep{2025AJ....169..170R, 2025arXiv251002639M}.
With both the Galactic prior evaluated by \texttt{gapmoe} and the light-curve likelihood made differentiable, gradient-based samplers such as Hamiltonian Monte Carlo can be applied directly to the full posterior.
Such methods can be particularly advantageous when parallax and lens orbital motion are included simultaneously, because the resulting inference involves a high-dimensional and strongly correlated parameter space.
This provides a promising route toward scalable inference for large samples of events while consistently retaining higher-order effects.
The same computational advantages would also facilitate the population-level inference described in Section~\ref{sec:population_inference}, which requires likelihood information from large numbers of events to be combined consistently.

The framework could also be extended to accommodate a broader range of Galactic models. Although the current implementation of \texttt{gapmoe} is based on the Koshimoto Galactic model, the same approach can be applied to other population models as long as their joint prior density for the relevant physical parameters can be represented in a differentiable manner. For Galactic models available primarily through forward simulations, flexible density estimators such as normalizing flows \citep{nflow} could provide a practical way to construct such differentiable representations while capturing their high-dimensional correlated distributions.

At the same time, our framework does not eliminate all sources of uncertainty.
While \texttt{gapmoe} provides a flexible model for the prior, systematic effects in the photometric data can still lead to a misspecified likelihood.
A complementary improvement is therefore to infer the photometric-noise model together with the microlensing parameters.
For example, Equation~\eqref{eq:likelihood_normal} can be generalized from independent errors to a Gaussian likelihood with covariance matrix $\bm{C}$,
\begin{align}
    \mathcal{L}(\bm{x})
    =
    {1 \over \sqrt{|2\pi\bm{C}|}}
    \exp\left[
    -\frac{1}{2}
    (\bm{f}-\bm{m}(\bm{x}))^{\mathsf{T}}\bm{C}^{-1}(\bm{f}-\bm{m}(\bm{x}))
    \right].
\end{align}
Common photometric-error renormalization schemes \citep{yee12} can be viewed as parametric models for the diagonal elements of $\bm{C}$, for example by fitting an overall error scale and/or an additional error floor.
More generally, models such as Gaussian processes \citep[e.g.,][]{gp,li19} can also parameterize the off-diagonal elements and thereby account for time-correlated residuals.
Jointly inferring these noise parameters with the microlensing parameters would complement the physically motivated prior provided by \texttt{gapmoe} by making the likelihood itself more flexible.

Taken together, these developments position our method as a powerful tool for the next generation of microlensing science.

\software{
\texttt{gapmoe}, \texttt{JAX} \citep{jax18},  \texttt{VBMicroLensing} \citep{boz10, boz18, boz21},
\texttt{corner} \citep{corner}, \texttt{emcee} \citep{emcee}
}



\begin{acknowledgments}
We thank Shota Miyazaki for carefully reading the manuscript and providing helpful comments, and the anonymous referee for constructive comments that improved the manuscript.
We also thank Zhecheng Hu, Wei Zhu, Subo Dong, and members of the infrared astronomy group at Osaka University for helpful conversations.
KN was supported by the JST Next Generation Researcher Challenging Research Program at Osaka University. Work by KM was supported by JSPS KAKENHI grant Nos.~21H04998, 25K07387, 26K22393, and 26H02074.
The authors used ChatGPT (OpenAI) to assist with language editing and improving the clarity of the manuscript.
The authors reviewed and edited all AI-assisted text and take full responsibility for the content of the manuscript.
\end{acknowledgments}

\appendix
\section{Observable-Physical Parameter Relation with Higher-Order Effects}\label{sec:parameterization}
In this appendix, we summarize the standard parameterization of binary-lens microlensing events with higher-order effects and provide the explicit relations between the observable light-curve parameters and the physical parameters of the system.

\subsection{Light-Curve Parameters}\label{ssec:parameterization_x}

The standard binary-lens microlensing model without higher-order effects is described by the following six parameters:
\begin{align}
t_0 &\quad \text{:time of closest approach between the source and the center of mass of the lens system} \notag\\
u_0 &\quad \text{: impact parameter (in units of the Einstein radius),} \notag\\
t_{\rm E} &\quad \text{: Einstein radius crossing time,}\notag \\
q &\quad \text{: mass ratio between the planet and the total lens mass,} \notag\\
s &\quad \text{: projected separation between the lens components (in units of Einstein radius),} \notag\\
\alpha &\quad \text{: angle of source trajectory relative to the binary axis.}\notag
\end{align}
The Einstein radius crossing time $t_{\rm E}$ is determined by the lens mass $M_{\rm L}$,
the distances to the lens and source ($D_{\rm L}$ and $D_{\rm S}$),
and the relative proper motion $\bm{\mu}{\rm rel}$ between them, as
\begin{align}
t_{\rm E} &= \frac{\theta_{\rm E}(M_{\rm L}, D_{\rm L}, D_{\rm S})}{|\bm{\mu}_{\rm rel}|}, \label{eq:tE}
\end{align}
where, $\theta_{\rm E}$ is the angular Einstein radius, given by
\begin{align} \label{eq:thE}
\theta_{\rm E} = \sqrt{\kappa M_{\rm L} \left( \frac{1~\mathrm{AU}}{D_{\rm L}} - \frac{1~\mathrm{AU}}{D_{\rm S}} \right)}, \quad \kappa = 8.144~\mathrm{mas}~M_\odot^{-1},
\end{align}
and the corresponding physical Einstein radius is given by
\begin{align}
R_{\rm E} = D_{\rm L}\,\theta_{\rm E}.\label{eq:RE}
\end{align}

In addition to the standard microlensing model, three higher-order effects are typically considered when aiming to constrain the physical parameters of the lens system: finite-source effects, microlensing parallax, and lens orbital motion.

\paragraph{Finite-source effect}
The finite-source effect introduces an additional parameter, the angular source radius in units of the angular Einstein radius,
\begin{align}
\rho &\equiv \frac{\theta_*}{\theta_{\rm E}}, \label{eq:rho}
\end{align}
where $\theta_*$ is the angular radius of the source star.  
Since $\theta_*$ can be estimated from the color and magnitude of the source, the measurement of $\rho$ effectively allows a determination of $\theta_{\rm E}$.

\paragraph{Parallax}
Microlensing parallax is introduced via the vector parameter $\bm{\pi}_{\rm E}$\citep{gou92,gou00,gou04}, which is defined as
\begin{align}
\bm{\pi}_{\rm E} &= \frac{\pi_{\rm rel}}{\theta_{\rm E}} \frac{\bm{\mu}_{\rm rel}}{|\bm{\mu}_{\rm rel}|}, \label{eq:piE}
\end{align}
where $\pi_{\rm rel} = \mathrm{AU}\left( D_{\rm L}^{-1} - D_{\rm S}^{-1} \right)$ is the lens-source relative parallax.

In addition to the parallax vector $\bm{\pi}_{\rm E}$, introducing the parallax effect requires specifying a reference time $t_{\rm ref}$.
The parallax is defined as the deviation of the observer's position from a frame centered on the Earth at $t_{\rm ref}$.
The choice of $t_{\rm ref}$ is arbitrary; for instance, in the \texttt{VBMicrolensing} \citep{boz10, boz21} implementation, $t_{\rm ref}$ is set to $t_0$ by default.
The same reference used is also used to define the parameters of the LOM introduced in the next.

\paragraph{Lens orbital motion}
The treatment of lens orbital motion depends on the level of complexity considered.  
For circular orbital motion, an additional three-dimensional vector $\boldsymbol{\gamma} = (\gamma_1, \gamma_2, \gamma_3)$ is introduced, defined as \citep{sko11}:
\begin{align}
\gamma_1 &\equiv \frac{1}{s} \frac{ds}{dt}, \quad \text{(expansion/contraction along the binary axis)} \notag\\
\gamma_2 &\equiv \frac{d\alpha}{dt}, \quad \text{(rotation in the sky plane)} \notag\\
\gamma_3 &\equiv \frac{1}{s} \frac{ds_z}{dt}, \quad \text{(motion along the line of sight)}\notag.
\end{align}
These components are expressed in units of $\mathrm{day}^{-1}$ and evaluated at $t = t_{\rm ref}$.
For describing eccentric orbits, two additional parameters are introduced \citep{boz21}:
\begin{align}
r_s &\equiv \frac{s_z}{s}, \quad \text{(line-of-sight separation normalized to projected separation)} \notag\\
a_s &\equiv \frac{a}{\sqrt{s^2 + s_z^2}}. \quad \text{(semi-major axis normalized to 3D separation)}\notag.
\end{align}

Given these light-curve parameters, the magnification of the source at any time $A(t)$ can be computed.  
The observed flux $f(t)$ is then modeled as  
\begin{align}
f(t) = f_{\rm s} A(t) + f_{\rm b},\label{eq:fs_fb}
\end{align}
where $f_{\rm s}$ and $f_{\rm b}$ are the source and blended fluxes, respectively, which are typically defined separately for each observatory and photometric band.  
Because $f_{\rm s}$ and $f_{\rm b}$ are linear parameters, their maximum-likelihood estimates can be determined analytically under the assumption that the observed fluxes $f_i$ follow independent Gaussian distributions---equivalent to a linear regression problem.
In our simulations, we adopt this analytic solution.

\subsection{Transformation between light-curve parameters and physical parameters}
\begin{figure}[t]
    \centering
    \includegraphics[width=0.5\textwidth]{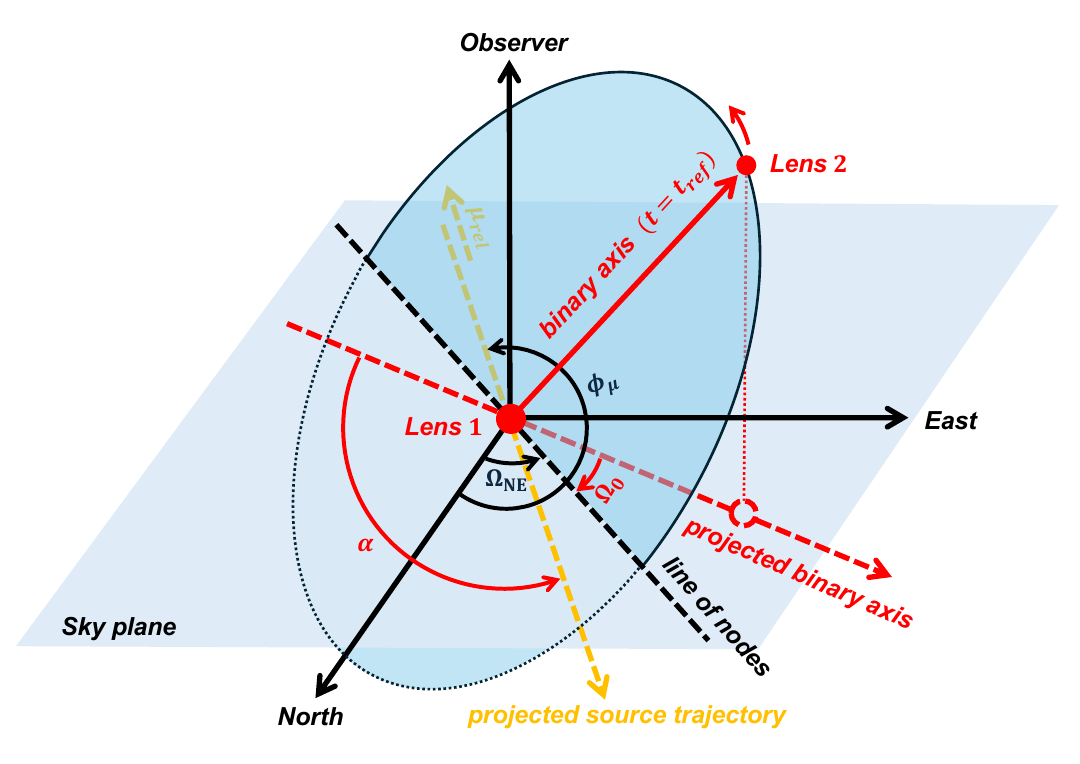}
    \includegraphics[width=0.4\textwidth]{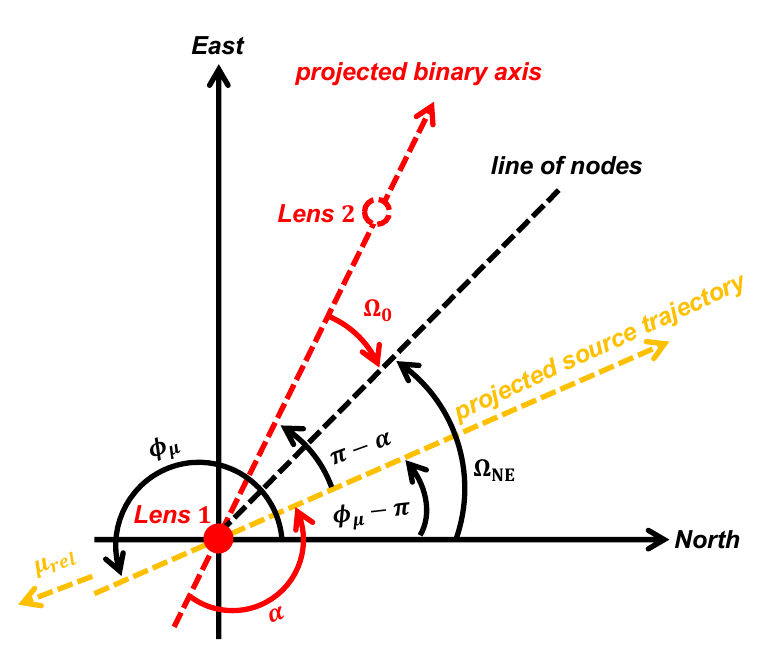}
    \caption{
    Geometry of the lens system and projected source trajectory in microlensing.  
    The left panel shows the three-dimensional configuration and the right panel its two-dimensional projection on the sky plane.  
    All angles appear in this figure are defined within the sky plane.
    Note that $\phi_\mu$ is defined by the direction of $\bm{\mu_{\rm rel}}$, taken from the source to the lens.  
    }
    \label{fig:orbit}
\end{figure}

In Section~\ref{subsec:params}, we introduced the full set of light-curve parameters $\bm{x}$, including higher-order effects, and the corresponding set of physical parameters $\bm{y}$. For convenience, we repeat their definitions here:
\begin{align}
\bm{x} &= (t_0, u_0, q, \alpha, t_{\rm E}, \rho, \bm{\pi}_{\rm E}, s, \gamma_1, \gamma_2, \gamma_3, r_s, a_s),\\
\bm{y} &= (t_0, u_0, q, M_{\rm L}, D_{\rm L}, D_{\rm S}, \bm{\mu}_{\rm rel}, a, e, i, \Omega_{\rm NE}, \omega, \nu).
\end{align}
Here, $a$ is the semi-major axis, $e$ is the eccentricity, $i$ is the inclination, $\Omega_{\rm NE}$ is the longitude of the ascending node in the North---East sky coordinate system, $\omega$ is the argument of periapsis, and $\nu$ is the true anomaly at the reference time.

With these parameter sets in hand, the next step is to establish an explicit transformation between $\bm{x}$ and $\bm{y}$. 
This mapping is required to incorporate Galactic priors defined in the physical parameter space into the inference carried out in $\bm{x}$. 
In this section, we therefore summarize the transformation between the microlensing observables and the corresponding physical parameters.

\paragraph{Expressions for ($M_{\rm L}$, $D_{\rm L}$, $D_{\rm S}$, $\bm{\mu}_{\rm rel}$)} From Eqs.~\eqref{eq:tE}, \eqref{eq:rho}, and \eqref{eq:piE}, the relative proper motion can be written as
\begin{align}
\bm{\mu}_{\rm rel} = \frac{\theta_*}{\rho\,t_{\rm E}} \cdot \frac{\bm{\pi}_{\rm E}}{|\bm{\pi}_{\rm E}|}.\label{eq:murel}
\end{align}
The angular source radius $\theta_*$ is typically estimated from the color and magnitude of the source star.  
When photometric data are available in both the $I$ and $V$ bands, the source magnitudes $I_{\rm S}$ and $V_{\rm S}$ can be derived from the source fluxes $f_{\rm s,I}$ and $f_{\rm s,V}$ obtained through the linear regression described in Equation~\eqref{eq:fs_fb}.  
Then, following \citet{boy14}, $\theta_*$ is often determined using the empirical relation:
\begin{align}
\log (2\theta_*) = 0.5014 + 0.4197 (V - I)_{\rm S,0} - 0.2 I_{\rm S,0}. \label{eq:theta_s}
\end{align}
In the simulations presented in this work, we adopt this relation to compute $\theta_*$.

Using Eqs.~\eqref{eq:thE} and \eqref{eq:piE} for $\theta_{\rm E}$, the lens mass and distance are given by
\begin{align}
M_{\rm L} &= \frac{\theta_{\rm E}}{\kappa |\bm{\pi}_{\rm E}|}, \label{eq:ML} \\
D_{\rm L} &= 1\mathrm{AU}~\left( |\bm{\pi}_{\rm E}| \theta_{\rm E} + \frac{1\mathrm{AU}}{D_{\rm S}} \right)^{-1}. \label{eq:DL}
\end{align}
Notice that this expression for $D_{\rm L}$ still depends on the unknown source distance $D_{\rm S}$. This can be determined by incorporating LOM through Kepler’s third law.  
The parameters $(\bm{\gamma}, a_s, r_s)$ defined in \citet{boz21} provide independent constraints on the semi-major axis and mean motion as:
\begin{align}
a &= R_{\rm E} a_s s \sqrt{1 + r_s^2}, \\
n &= \frac{|\bm{\gamma}|}{a_s \sqrt{(2a_s - 1)(1 + r_s^2)}}.
\end{align}
From Kepler’s third law,
\begin{align}
n = \sqrt{ \frac{G M_{\rm L}}{a^3} },
\end{align}
and substituting the expression for $a$ into this, we obtain
\begin{align}
\frac{G M_{\rm L}}{R_{\rm E}^3} = s^3 a_s \sqrt{1 + r_s^2} \cdot \frac{ \gamma_1^2 + \gamma_2^2 + \gamma_3^2 }{2a_s - 1}.
\end{align}
Substituting Equation~\eqref{eq:RE} for $R_{\rm E}$ and solving for $D_{\rm S}$, we arrive at
\begin{align}
D_{\rm S} = \frac{1~{\rm AU}}{\theta_{\rm E}} \left( \left[ \frac{(1~{\rm AU})^3}{G M_{\rm L}} s^3 a_s \sqrt{1 + r_s^2} \cdot \frac{ \gamma_1^2 + \gamma_2^2 + \gamma_3^2 }{2a_s - 1} \right]^{1/3} - \pi_{\rm E} \right)^{-1}. \label{eq:DS}
\end{align}
Once $D_{\rm S}$ is known, $D_{\rm L}$ can be immediately determined via Equation~\eqref{eq:DL}.

\paragraph{Expressions for binary orbital elemetns}
We define an orthonormal triad $\{\hat{\bm{x}}_0,\hat{\bm{y}}_0,\hat{\bm{z}}_0\}$ at the reference epoch $t_{\rm ref}$ such that
\begin{align}
&\text{(i) $\hat{\bm{x}}_0$ is aligned with the projected binary axis (from the primary to the secondary),}\notag\\
&\text{(ii) $\hat{\bm{z}}_0$ points toward the observer, and}\notag\\
&\text{(iii) $\hat{\bm{y}}_0 \equiv \hat{\bm{z}}_0 \times \hat{\bm{x}}_0$ completes a right--handed system.}\notag
\end{align}
In this basis, and following the definitions of $\bm{\gamma}$ and $r_s$, the relative position and velocity of the secondary with respect to the primary at $t_{\rm ref}$ are \citep{boz21}:
\begin{align}
\bm{v} &= R_{\rm E}\,s\left(\gamma_1\,\hat{\bm{x}}_0 + \gamma_2\,\hat{\bm{y}}_0 + \gamma_3\,\hat{\bm{z}}_0\right), \label{eq:v_lincomb}\\
\bm{r} &= R_{\rm E}\,s\left(\hat{\bm{x}}_0 + r_s\,\hat{\bm{z}}_0\right). \label{eq:r_lincomb}
\end{align}

They fix the specific angular momentum vector $\bm{h}$ and the Laplace--Runge--Lenz vector $\bm{e}$ as
\begin{align}
\bm{h} = \bm{r} \times \bm{v}, \quad
\bm{e} = \frac{\bm{v} \times \bm{h}}{G M_{\rm L}} - \frac{\bm{r}}{|\bm{r}|},
\end{align}
where eccentricity $e$ is given by $|\bm{e}|$.

To describe the orientation of the orbit, we define an new orthonormal triad:
\begin{align}
\hat{\bm{x}} = \frac{\bm{e}}{|\bm{e}|}, \quad
\hat{\bm{z}} = \frac{\bm{h}}{|\bm{h}|}, \quad
\hat{\bm{y}} = \hat{\bm{z}} \times \hat{\bm{x}}. \label{eq:xyz}
\end{align}
Using these, the orbital angles are obtained as follows.
The true anomaly $\nu$ is given by
\begin{align}
\cos \nu = \frac{\bm{r}}{|\bm{r}|} \cdot \hat{\bm{x}}, \quad
\sin \nu = \frac{\bm{r}}{|\bm{r}|} \cdot \hat{\bm{y}}.
\end{align}
The inclination $i$, longitude of ascending node $\Omega_0$, and argument of periapsis $\omega$ are
\begin{align}
\cos i = \hat{\bm{z}}\cdot{\hat{\bm{z}_0}}, \quad
\sin \Omega_0 = \frac{\hat{\bm{z}}\cdot{\hat{\bm{x}_0}}}{\sin i}, \quad
\cos \Omega_0 = -\frac{\hat{\bm{z}}\cdot{\hat{\bm{y}_0}}}{\sin i}, \quad
\sin \omega = \frac{\hat{\bm{x}}\cdot\hat{\bm{z}_0}}{\sin i}, \quad
\cos \omega = \frac{\hat{\bm{y}}\cdot\hat{\bm{z}_0}}{\sin i}.
\end{align}
Here, the subscript “0” in $\Omega_0$ denotes that the angle is measured 
in the reference frame $\{\hat{\bm{x}}_0, \hat{\bm{y}}_0, \hat{\bm{z}}_0\}$ at $t_{\rm ref}$, 
whereas $i$, $\omega$, and $\nu$ are defined within the orbital plane and are frame independent.
The longitude of ascending node measured in the North--East sky frame, $\Omega_{\rm NE}$, is given by
\begin{align}
\Omega_{\rm NE} = \Omega_0 + \phi_\mu - \alpha, \label{eq:Omega_NE}
\end{align}
where $\phi_\mu$ is the position angle of the relative proper motion vector $\bm{\mu}_{\rm rel}$, 
measured east of north from the source to the lens (see Fig.~\ref{fig:orbit}).

\bibliography{reference}

\end{document}